\documentclass[a4paper,11pt]{article}

\pdfoutput=1

\usepackage{jcappub}

\usepackage[T1]{fontenc}
\usepackage{subfig}
\usepackage{booktabs}
\subheader{DESY 20-203}
\title{\boldmath The Effective Field Theory and Perturbative Analysis for Log-Density Fields}

\author[a]{Henrique Rubira}
\author[b, c]{and Rodrigo Voivodic}

\affiliation[a]{Deutsches Elektronen-Synchrotron DESY, Notkestra\ss e 85, D-22607 Hamburg, Germany}
\affiliation[b]{ Departamento de F\'{\i}sica Matem\'atica, Instituto de F\'{\i}sica, Universidade de S\~ao Paulo,\\ R.  do  Mat\~ao  1371,  05508-090,  S\~ao Paulo, SP, Brazil}
\affiliation[c]{\small Max-Planck-Institut f\"ur Astrophysik, Karl-Schwarzschild-Stra\ss e~1, 85748 Garching, Germany}

\emailAdd{henrique.rubira@desy.de}
\emailAdd{rodrigo.voivodic@usp.br}

\abstract{A logarithm transformation over the matter overdensity field $\delta$ brings information from the bispectrum and higher-order n-point functions to the power spectrum. We calculate the power spectrum for the log-transformed field $A$ at one, two and three loops using perturbation theory (PT). We compare the results to simulated data and give evidence that the PT series is asymptotic already on large scales, where the $k$ modes no longer decouple. This motivates us to build an alternative perturbative series for the log-transformed field that is not constructed on top of perturbations of $\delta$ but directly over the equations of motion for $A$ itself. This new approach converges faster and better reproduces the large scales at low $z$. We then show that the large-scale behaviour for the log-transformed field power spectrum can be captured by a small number of free parameters. Finally, we add the counter-terms expected within the effective field theory framework and show that the theoretical model, together with the IR-resummation procedure, agrees with the measured spectrum with percent precision until $k \simeq 0.38 $ Mpc$^{-1}$h at $z=0$. It indicates that the non-linear transformation indeed linearizes the density field and, in principle, allows us to access information contained on smaller scales.}

\begin{document}
\maketitle
\flushbottom

\section{Introduction}
\label{sec:introduction}

The effective field theory of large scale structure (EFTofLSS) program \cite{Baumann:2010tm, Carrasco:2012cv, Carrasco:2013mua, Konstandin:2019bay} made possible to constrain cosmology within percent precision. Recent astonishing achievements include applying EFTofLSS to survey data \cite{DAmico:2019fhj,Ivanov:2019hqk,Colas:2019ret, Philcox:2020vvt} and to a blinded cosmology challenge \cite{Nishimichi:2020tvu}. The EFT is constructed on top of the so-called perturbation theory (PT) \cite{Bernardeau:2001qr} for the matter overdensity $\delta$, defined according to $\rho(\mathbf{x},\tau) = \bar{\rho}(\tau)[1+\delta(\mathbf{x},\tau)]$, where $\bar{\rho}(\tau)$ is the mean density at the background and $\tau$ is the conformal time.

\begin{figure}[ht]
\centering
  \includegraphics[width=0.48\textwidth]{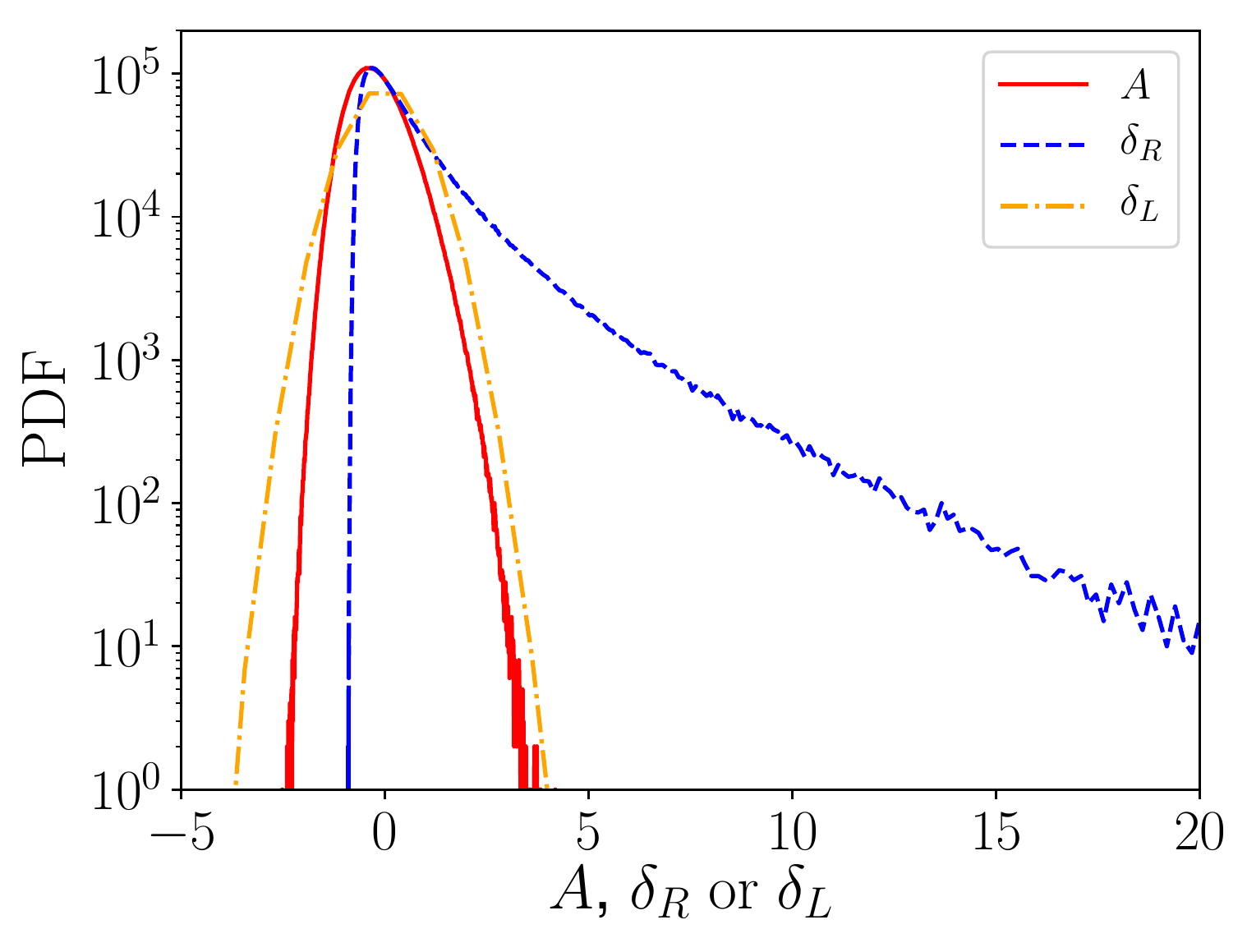}
  \includegraphics[width=0.49\textwidth]{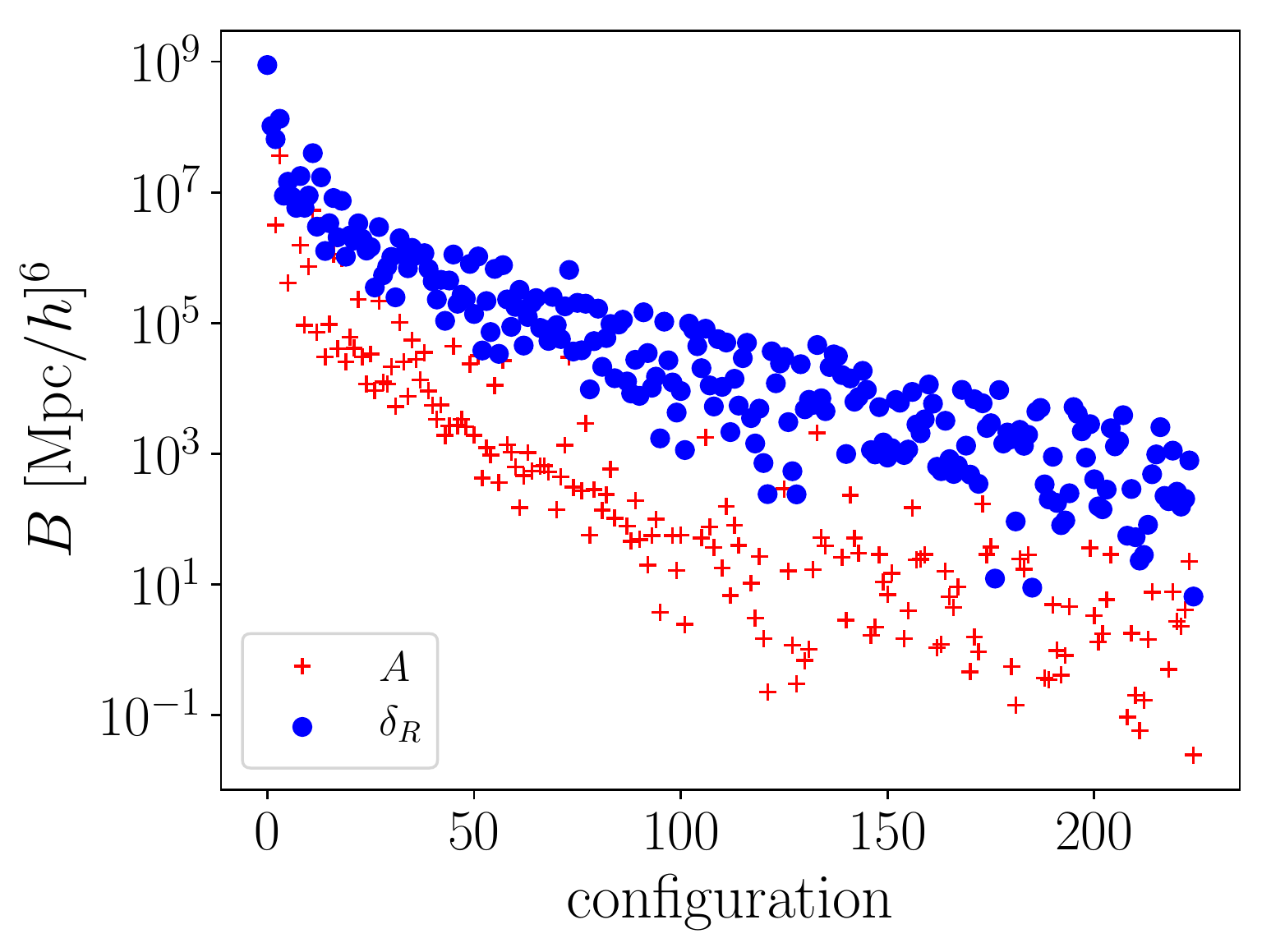}
\caption{\label{fig:bispecpdf}%
\small On the left, the probability density function for $A$ (red solid), $\delta$ (blue dashed) and the linearly extrapolated initial ($z=99$) density $\delta_L$ (orange dot-dashed) at $z = 0$. We smoothed $\delta$ using a Gaussian filter with a smoothing scale of $4$ Mpc h$^{-1}$. On the right, the bispectrum for all triangular configurations. The comparison between the bispectra amplitude makes evident the Gaussianization process of the logarithm transformation. The right tail of the PDFs makes clear the linearization property of this transformation.
}
\end{figure}

An important aspect of the overdensity evolution is that it is non-linear, making $\delta$ deviate from an initially Gaussian random distribution. 
As shown by \cite{Coles:1991if}, the lognormal distribution arises by considering the velocity to scale linearly in conformal time\footnote{Even though the lognormal approximation might capture some information about the displacement field, it is not strictly speaking the same as the Zel'dovich approximation \cite{Zeldovich:1969sb}. Zel'dovich considers the velocity to be proportional to the linear density and, for instance, predicts the formation of caustics that are not present in the lognormal field \cite{Coles:1991if}. }. 
One can therefore construct a Gaussian-like distribution \cite{weinberg,Szapudi:2002cr} by performing a non-linear transformation on $\delta$ as  
\begin{equation} \label{eq:Adef}
    A (\mathbf{x},\tau)= \ln{\left[ 1+\delta_R(\mathbf{x},\tau)\right]}\,,
\end{equation}
where $\delta_R$ is the overdensity field smoothed by a filter on a scale $R$. The Gaussianization can be seen on the left panel of Fig.~\ref{fig:bispecpdf}, where we compare the PDFs for $A$ and $\delta_R$. 

A central point to compare is how physical information is distributed along the moments and n-point functions of $A$ and $\delta$  \cite{Carron:2011et,Carron:2012hk}. Whilst for a random almost Gaussian field ($A$-like) most of the information is encoded in its first two moments, for its exponential counterpart ($\delta$-like) the information is distributed along all moments or even lost among them\footnote{The fact that information can be lost in the moments expansion is a consequence of the fact that the projection of the logarithm field into a polynomial basis fails when the variance of the field gets larger \cite{Carron:2011et}.}. The field  Gaussianization can also bring information from higher-order n-point functions down to lower ones (e.g. information from the bispectrum of $\delta$ will be encoded in the power spectrum of $A$). 
We display on the right panel of Fig.~\ref{fig:bispecpdf} the bispectrum for $A$ and $\delta$. The Gaussianization feature present in the non-linear transformation~\eqref{eq:Adef} becomes clear when comparing the amplitude of the bispectrum for $A$ and $\delta_R$.

\begin{figure}[ht]
\centering
  \includegraphics[width=0.35\textwidth]{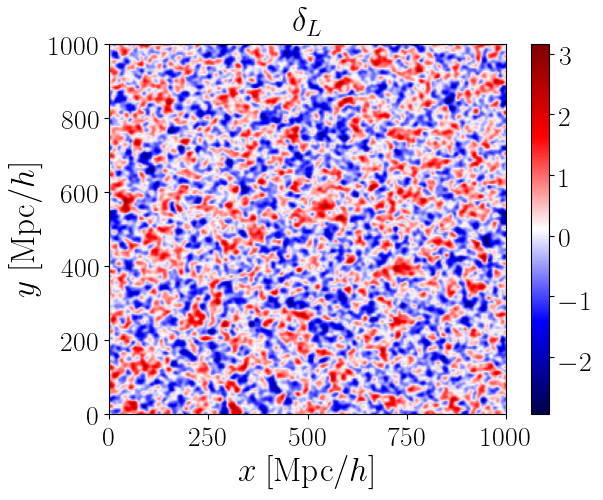}
  \includegraphics[width=0.305\textwidth]{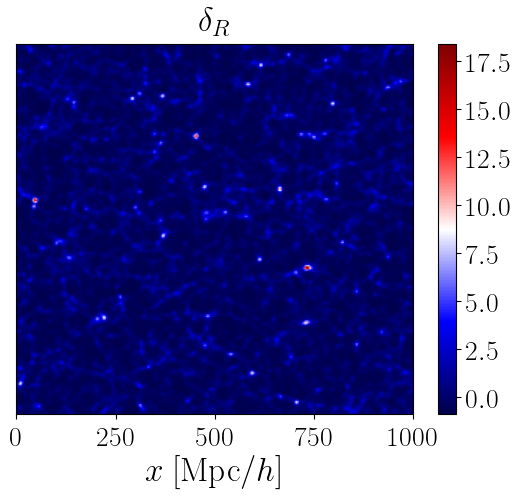}
  \includegraphics[width=0.305\textwidth]{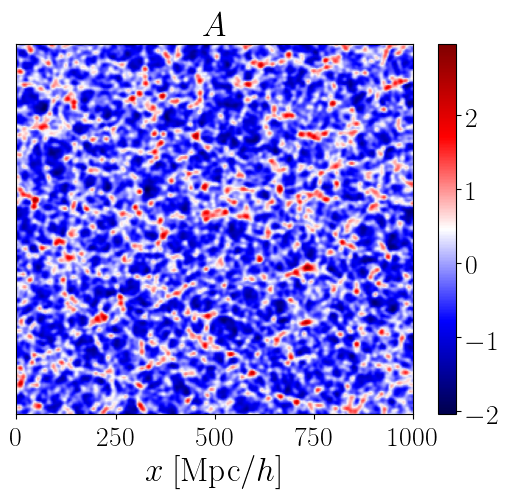}
\caption{\label{fig:densities}%
\small  A density slice of the N-body simulation at redshift zero. In the left, the initial condition from $z=99$ rescaled by the linear growth function. In the middle the density field $\delta_{R}$ and in the right the field $A = \ln{\left[ 1+\delta_R\right]}$. For all the cases, we used $R = 4$ Mpc h$^{-1}$ with an exponential filter. 
}
\end{figure}

In parallel to the Gaussianization effect induced by the transformation~(\ref{eq:Adef}), another central point to be analysed is the process of linearization of the field. Ultimately, non-linearities are induced by couplings between modes. Once the equations of motion for $A$ are different than for $\delta$ \cite{Wang:2011fj}, one expects both basis to describe non-linearities in a different way. Within the perturbation theory language, the vertices that describe the couplings for the $A$ modes are different than those for $\delta$. Besides that, $\delta$ assumes values much larger than $1$ at $z=0$, while the typical values for $A$ barely get larger than $1$ (see Figs.~\ref{fig:bispecpdf} and \ref{fig:densities}). 

Concisely, we have for the field $A$ compared to the field $\delta$ that: 
\begin{itemize}
    \item it is more Gaussian, implying that the higher n-point functions are smaller. Consequently, more information is available in the two-point function once the connected four-point function is smaller;
    \item it is expected to be more linear, implying in a faster convergence for the same value of $k$.
\end{itemize} 

Those results were described in the series of papers \cite{Neyrinck:2009fs,Neyrinck:2010th,Neyrinck:2011xm} and in \cite{Wang:2011fj}. Analysing the covariance matrix for the $A$ power spectrum, they have shown that the signal extracted out of the non-linear $k$ modes is larger for $A$ than for $\delta$ (see also \cite{Rimes:2005dz, Lee:2008qy}). Another important result that motivates to work with the log transformation concerns the position of the BAO peak \cite{McCullagh:2012nu}. While the BAO peak for $\delta$ shifts when  Zel'dovich corrections are taken into account, this shift is much smaller for $A$ due to bulk flow motions.

Recently the so-called marked transformation, another kind of non-linear transformation of $\delta$, has gathered a lot of attention. It was shown by \cite{Massara:2020pli} that marked fields can improve the constraints in the neutrino mass sum by a factor of 80. An effective field theory for marked fields was constructed by \cite{Philcox:2020fqx,Philcox:2020srd}. 
Once $A$ is also more linear and increases the weight of cosmic voids in the correlation functions, we also expect it to provide more stringent constraints on neutrinos, dark energy, and gravity \cite{Kreisch, Voivodic1, Perico}. 

We revisit in this work the idea of constructing a perturbative prediction for the  power spectrum of the log field motivated by the idea that non-linear transformations of the overdensity field can lead to better constraints on the cosmological parameters. 
We first analyse the convergence of the perturbative series for $A$. For this, we Taylor expand the logarithm in Eq.~(\ref{eq:Adef}) and extend the one-loop analysis of \cite{Wang:2011fj} up to three loops. We label this approach as Taylor-$A$PT. Comparing it to N-body simulations, we give strong evidence that the series constructed by that manner does not converge to the expected simulation result at low redshift even on large scales\footnote{Notice that one can determine a convergence radius for $\delta$ \cite{Konstandin:2019bay}, while for the $A$ case it already fails on the large scales. As will be shown later in the text, the $k$ modes no longer decouple for $A$ in the limit $k \rightarrow 0$.}. It is a consequence of the lack of convergence of the PT series for $\delta$ on intermediate scales, already pointed out by \cite{Blas:2013aba, Pajer:2017ulp, Konstandin:2019bay}. We then propose an alternative perturbative approach for $A$ directly through the equations of motion, which we label as EoM-$A$PT. The series constructed on that way has a better loop-hierarchy and gets closer to the simulation result at low redshift. Afterwards, we show that the inclusion of a few free parameters, already considered in \cite{Philcox:2020srd}, can improve the small $k$ behaviour for both schemes.

Next, we indicate the counter-terms needed by the effective field theory for $A$ at one and two-loop order. We show that the UV dependence of the loop integrals for the $A$ field demands us to add a new counter-term at two loops compared to the EFT for $\delta$. This new counter-terms comes from the bispectrum for $\delta$ and is a payback from bringing information from the bispectrum down to the power spectrum.  We also perform a IR-resummation scheme based on \cite{Senatore:2014via,Cataneo:2016suz}.

We compare our theoretical results to a suite of 26 N-body simulation ran with the \texttt{RAMSES} code \cite{Teyssier}. We use a box with $1024$ Mpc h$^{-1}$ and $512^{3}$ particles. The cosmological parameters used are ($\Omega _{m}$, $\Omega _{b}$, $\Omega _{\Lambda}$, $\sigma _{8}$, $n_{s}$, $H_{0}$) = (0.307, 0.048, 0.693, 0.829, 0.96, 67.8). We calculate the density grid using a Gaussian filter
\begin{equation}
    W(r/R) = e^{-\frac{1}{2}\frac{r^2}{R^2}},
    \label{eq:window}
\end{equation}
with $R = 4$ Mpc h$^{-1}$. Notice that the field excursions for $\delta$ in \cite{Neyrinck:2009fs} are larger, since they use a top hat filter and a smaller $R$. We picked this radius value because its reciprocal $k$ ($k_R \sim \sqrt{2} \times 0.25 $ Mpc$^{-1}$h) is above the typical edge of the one-loop perturbative prediction for $\delta$ and, as it will be seen in Sec.~\ref{sec:largescales}, the series for $A$ is still perturbative. Besides that, using a Gaussian filter leads to a smoother connection between the measurements in position space and the theory in Fourier space. 

The structure of this work is the following. We start Sec.~\ref{sec:theory} by settling the theoretical basis for both the Taylor-$A$PT and EoM-$A$PT schemes. We discuss the convergence of the series in the same section, which motivates us to introduce a set of free-parameters and counter-terms.  Sec.~\ref{sec:results} is dedicated to compare the EFT theory and the simulated data. We also compare the  data to two other schemes: a power-law fitting function based on the work \cite{Repp:2016gxt} and the large-scale bias expansion model \cite{Desjacques:2016bnm}. We conclude in Sec.~\ref{sec:conclusion}. In App.~\ref{app:scaling}, we compute the scaling of the PT integrals up to two loops.

\section{Theoretical Models}
\label{sec:theory}

In this section we present the theoretical structure underlying the perturbation theory framework for the logarithm-transformed field. We start by presenting the Taylor expansion of the log field in terms of $\delta_R$. We show that within this scheme the PT series does not converge to the N-body simulation result on the largest scales, which motivates us to consider an alternative perturbative approach directly through the equations of motion. 

Next, we comment on how the smoothing scale $R$ affects the large scales and whether perturbation theory can capture this effect by adding a couple of free parameters. Finally, we comment on the effective field theory approach for $A$.  

\subsection{Taylor-$A$PT} \label{sec:deltapt}

We now construct a perturbative series for the $A$ field by expanding the $\log$ in Eq.~(\ref{eq:Adef}) in terms of $\delta_R$. We label this expansion as Taylor-$A$PT. This method was already described by \cite{Wang:2011fj} at one-loop level and here we extend it to two loops. We also calculate the three-loop contribution on the large scales, aiming to study the series behaviour on this limit.

We first expand the $\log$  as
 \begin{eqnarray}\label{eq:TaylorA}
    A (\mathbf{x}, \tau) \equiv \ln{[1+\delta_R(\mathbf{x}, \tau)]} &=& \sum_{n=1}^{\infty} (-1)^{n+1}\frac{\delta_R^n}{n}  
    \\  &=& \delta_R - \frac{1}{2}\delta_R^2 + \frac{1}{3}\delta_R^3 - \frac{1}{4}\delta_R^4  + \frac{1}{5}\delta_R^5 - \frac{1}{6}\delta_R^6 + \frac{1}{7}\delta_R^7 + \dots  \nonumber
\end{eqnarray}
 The temporal part decouples from the spacial one in an EdS universe and we define the perturbative expansion for $A$ analogously to what is done for $\delta$
\begin{eqnarray}\label{eq:deltapt_pert}
A (\mathbf{x}, \tau) = \sum_n a^n(\tau)\, A^{(n)}(\mathbf{x})\,, \qquad \theta = -\mathcal{H}(\tau)\sum_n a^n(\tau)\, \theta^{(n)}(\mathbf{x})\,,
\end{eqnarray}
where $\mathcal{H}$ is the comoving Hubble rate ($\mathcal{H} = da/d\tau$). 

For cosmologies different than EdS, the time evolution is well described by the approximation $a^n(\tau) \rightarrow D ^{n}(\tau)$ \cite{Garny:2020ilv,Steele:2020tak}. In Fourier space, each expansion term in Eq.~(\ref{eq:deltapt_pert}) is given by
\small{
\begin{eqnarray}
    A^{(1)} (\mathbf{k}) &=& \delta_R^{(1)}(\mathbf{k}) \label{eq:A1}\,,
    \\
    A^{(2)} (\mathbf{k}) &=& \delta_R^{(2)}(\mathbf{k}) - \left[ \frac{1}{2}\delta_R^{(1)} \ast \delta_R^{(1)} \right] (\mathbf{k}) \,,
    \\
    A^{(3)} (\mathbf{k}) &=& \delta_R^{(3)}(\mathbf{k}) - \left[\delta_R^{(1)} \ast \delta_R^{(2)}\right] (\mathbf{k}) + \left[ \frac{1}{3}\delta_R^{(1)} \ast \delta_R^{(1)}\ast \delta_R^{(1)}\right] (\mathbf{k}) \,,
    \\
    A^{(4)} (\mathbf{k}) &=& \delta_R^{(4)}(\mathbf{k}) - \left[\frac{1}{2} \delta_R^{(2)} \ast \delta_R^{(2)} \right] (\mathbf{k}) - \left[ \delta_R^{(1)} \ast \delta_R^{(3)}\right] (\mathbf{k}) 
    \\
    &+&  \left[\delta_R^{(1)} \ast \delta_R^{(1)}\ast \delta_R^{(2)}\right] (\mathbf{k}) - \left[  \frac{1}{4}\delta_R^{(1)} \ast \delta_R^{(1)}\ast \delta_R^{(1)}\ast \delta_R^{(1)} \right] (\mathbf{k}) \,, \nonumber
    \\
    A^{(5)} (\mathbf{k}) &=& \delta_R^{(5)}(\mathbf{k}) -  \left[ \delta_R^{(1)} \ast \delta_R^{(4)}\right] (\mathbf{k})  - \left[ \delta_R^{(2)} \ast \delta_R^{(3)}\right] (\mathbf{k}) +  \left[ \delta_R^{(1)} \ast \delta_R^{(1)}\ast \delta_R^{(3)}\right] (\mathbf{k}) \label{eq:A5} \\ &+&   \left[ \delta_R^{(1)} \ast \delta_R^{(2)}\ast \delta_R^{(2)}\right] (\mathbf{k})  
    -  \left[ \delta_R^{(1)} \ast \delta_R^{(1)}\ast \delta_R^{(1)}\ast \delta_R^{(2)}\right] (\mathbf{k}) \nonumber \\
    &+& \left[ \frac{1}{5}\delta_R^{(1)} \ast \delta_R^{(1)}\ast \delta_R^{(1)}\ast \delta_R^{(1)}\ast \delta_R^{(1)}\right] (\mathbf{k}) \nonumber \,, 
\end{eqnarray}}\normalsize
where $\ast$ stands for the convolution product. To avoid cluttering the text, we omit the sixth and seventh order terms present at three loops.

When considering the perturbation theory for $\delta$ and $\theta$, the $n^{th}$ order kernels are written as a convolution of $n$ linear fields with the kernels $F^{(n)}$ and $G^{(n)}$
\begin{eqnarray}
    \delta^{(n)}(\mathbf{k}) &=& \int_{\mathbf{q}_{1\dots n}}\delta_{\rm D}(\mathbf{q}_{1\dots n} - \mathbf{k})\, F^{(n)}(\mathbf{q}_1, \dots, \mathbf{q}_n)\,\delta^{(1)}(\mathbf{q}_1)\dots \delta^{(1)}(\mathbf{q}_n)  \, , \\
    \theta^{(n)}(\mathbf{k}) &=& \int_{\mathbf{q}_{1\dots n}}\delta_{\rm D}(\mathbf{q}_{1\dots n} - \mathbf{k})\, G^{(n)}(\mathbf{q}_1, \dots, \mathbf{q}_n)\,\delta^{(1)}(\mathbf{q}_1)\dots \delta^{(1)}(\mathbf{q}_n)  \, ,
\end{eqnarray}
where $F^{(n)}$ and $G^{(n)}$ are obtained through recursive relations \cite{Bernardeau:2001qr}. Analogously, we can write for $A$ 
\begin{eqnarray}
    A^{(n)}(\mathbf{k}) &=& \int_{\mathbf{q}_{1\dots n} }\delta_{\rm D}(\mathbf{q}_{1\dots n} - \mathbf{k})\, M^{(n)}(\mathbf{q}_1, \dots, \mathbf{q}_n)\,\delta^{(1)}(\mathbf{q}_1)\dots \delta^{(1)}(\mathbf{q}_n)  \, ,
\end{eqnarray}
as a function of the new kernels $M$. We kept the definition of the velocity field and thus the velocity kernels are unchanged. 
Using the notation $\int_{\mathbf{q}_{1\dots n}} = \int \frac{d^3 \mathbf{q}_1}{(2\pi)^3} \dots \int \frac{d^3 \mathbf{q}_n}{(2\pi)^3}$,
one can use Eqs.~(\ref{eq:A1})-(\ref{eq:A5}) to relate $M^{(n)}$ to $F^{(n)}$ 
\small{\begin{eqnarray}
    M^{(1)}(\mathbf{q}_1) &=& W(q_1 R)F^{(1)}(q_{1}) \,,\label{eq:M1}\\
    M^{(2)}(\mathbf{q}_1, \mathbf{q}_2) &=& W(q_{12} R)F^{(2)}(\mathbf{q}_1, \mathbf{q}_2) - \frac{1}{2}W(q_1 R)W(q_2 R) \,, \label{eq:M2}\\
    M^{(3)}(\mathbf{q}_1, \mathbf{q}_2, \mathbf{q}_3) &=& W(q_{123}R)F^{(3)}(\mathbf{q}_1, \mathbf{q}_2, \mathbf{q}_3) \label{eq:M3} \\
&-& \frac{1}{3}\left[W(q_{12} R)W(q_3 R)F^{(2)}(\mathbf{q}_1, \mathbf{q}_2) + 2\,\textrm{sym} \right] \nonumber \\ 
&+&  \frac{1}{3}W(q_1 R)W(q_2 R)W(q_3 R) \nonumber \,,  \\
    M^{(4)}(\mathbf{q}_1, \mathbf{q}_2, \mathbf{q}_3, \mathbf{q}_4) &=& W(q_{1234}R)F^{(4)}(\mathbf{q}_1, \mathbf{q}_2, \mathbf{q}_3, \mathbf{q}_4) \\
    &-& \frac{1}{2} \frac{1}{3}\left[ W(q_{12} R)W(q_{34} R)F^{(2)}(\mathbf{q}_1, \mathbf{q}_2)F^{(2)}(\mathbf{q}_3, \mathbf{q}_4) + 2\,\textrm{sym} \right] \nonumber \\ 
    &-& \frac{1}{4}\left[ W(q_{123}R)W(q_4 R)F^{(3)}(\mathbf{q}_1, \mathbf{q}_2, \mathbf{q}_3) + 3\,\textrm{sym}\right] \nonumber  \\
    &+&  \frac{1}{6}\left[ W(q_{12} R)W(q_{3} R)W(q_{4} R)F^{(2)}(\mathbf{q}_1, \mathbf{q}_2) + 5\,\textrm{sym} \right] \nonumber \\
    &-&  \frac{1}{4}W(q_1 R)W(q_2 R)W(q_3 R)W(q_4 R)\nonumber \,, \\
    M^{(5)}(\mathbf{q}_1, \mathbf{q}_2, \mathbf{q}_3, \mathbf{q}_4, \mathbf{q}_5) &=& W(q_{12345}R)F^{(5)}(\mathbf{q}_1, \mathbf{q}_2, \mathbf{q}_3, \mathbf{q}_4, \mathbf{q}_5)\label{eq:M5} \\
    &-&  \frac{1}{5}\left[ W(q_{1234}R)W(q_5 R)F^{(4)}(\mathbf{q}_1, \mathbf{q}_2, \mathbf{q}_3, \mathbf{q}_4) + 4\,\textrm{sym}\right] \nonumber  \\
    &-& \frac{1}{10}\left[ W(q_{12} R)W(q_{345}R)F^{(2)}(\mathbf{q}_1, \mathbf{q}_2)F^{(3)}(\mathbf{q}_3, \mathbf{q}_4, \mathbf{q}_5) + 9\,\textrm{sym}\right] \nonumber \\
    &+&  \frac{1}{10}\left[ W(q_{123}R)W(q_{4}R)W(q_{5}R)F^{(3)}(\mathbf{q}_1, \mathbf{q}_2, \mathbf{q}_3) + 9\,\textrm{sym} \right] \nonumber \\
    &+&  \frac{1}{15}\left[ W(q_{12} R)W(q_{34} R)W(q_{5} R)F^{(2)}(\mathbf{q}_1, \mathbf{q}_2)F^{(2)}(\mathbf{q}_3, \mathbf{q}_4) + 14\,\textrm{sym}  \right]  \nonumber \\
    &-&  \frac{1}{10}\left[ W(q_{12} R)W(q_{3} R)W(q_{4} R)W(q_{5} R)F^{(2)}(\mathbf{q}_1, \mathbf{q}_2) + 9\,\textrm{sym} \right] \nonumber \\
    &+& \frac{1}{5}W(q_1 R)W(q_2 R)W(q_3 R)W(q_4 R)W(q_5 R) \nonumber \,.
\end{eqnarray}}\normalsize

Defining the power spectrum of $A$ as 
\begin{equation}
    \left\langle A(\textbf{k}, \eta) A(\textbf{k}', \eta) \right\rangle = (2\pi)^{3}\delta_{\rm D}(\textbf{k} + \textbf{k}')P_{A}(k, \eta) \,,
\end{equation}
the one, two and three-loop predictions are given by
\begin{eqnarray}
    P_{A}^{PT, 1L}  &=& P_{A}^{0L} + P_{A}^{1L}  =  P_{A}^{(11)} + 2 P_{A}^{(13)} + P_{A}^{(22)}  \,, \label{eq:P1l_taylor}
    \\
    P_{A}^{PT, 2L} &=& P_{A}^{0L} + P_{A}^{1L} + P_{A}^{2L} 
    \\ 
    &=&  P_{A}^{\rm PT, 1L} + 2 P_{A}^{(15)} + 2 P_{A}^{(24)} + P_{A}^{(33a)} + P_{A}^{(33b)} \,.\nonumber
    \\
    P_{A}^{PT, 3L}  &=& P_{A}^{0L} + P_{A}^{1L} + P_{A}^{2L} + P_{A}^{3L} 
    \\ 
    &=&  P_{A}^{PT, 2L} + 2 P_{A}^{(17)} + 2 P_{A}^{(26)} + 2P_{A}^{(35a)} + 2P_{A}^{(35b)}+ P_{A}^{(44a)} + P_{A}^{(44b)} \,.\nonumber
\end{eqnarray}
We use $P_{A}^{\bullet L}$ to refer to each loop contribution and $P_{A}^{PT, \bullet L}$ states for the sum of all loop terms that contribute up to some order. Each loop diagram is calculated as the usual way for $\delta$ but replacing $F\rightarrow M$:
\small \begin{eqnarray}
    P_{A}^{(11)}(k) &=& \left[ M^{(1)}(k)\right]^2 P_L(k) \label{eq:PA_11}\,, \\
    P_{A}^{(13)}(k) &=& 3 M^{(1)}(k)P_L(k) \int_\mathbf{q} M^{(3)}(\mathbf{k},\mathbf{q},-\mathbf{q}) P_L(q)\,,\label{eq:P13} \\
    P_{A}^{(22)}(k) &=& 2 \int_\mathbf{q} \left[ M^{(2)}(\mathbf{k}-\mathbf{q},\mathbf{q})\right]^2 P_L(|\mathbf{k}-\mathbf{q}|)P_L(q)\,, \label{eq:P22}\\
    P_{A}^{(15)}(k) &=& 15 M^{(1)}(k)P_L(k) \int_{\mathbf{q}_{12}} M^{(5)}(\mathbf{k},\mathbf{q}_1,-\mathbf{q}_1,\mathbf{q}_2,-\mathbf{q}_2) P_L(q_1)P_L(q_2) \,, \\
    P_{A}^{(24)}(k) &=& 12 \int_{\mathbf{q}_{12}} M^{(2)}(\mathbf{k}-\mathbf{q}_1,\mathbf{q}_1) M^{(4)}(\mathbf{k}-\mathbf{q}_1,\mathbf{q}_1,\mathbf{q}_2,-\mathbf{q}_2) P_L(|\mathbf{k}-\mathbf{q}_1|)P_L(q_1)P_L(q_2) \,,\\    
    P_{A}^{(33a)}(k) &=& 9P_L(k) \int_{\mathbf{q}_{12}} M^{(3)}(\mathbf{k},\mathbf{q}_1,-\mathbf{q}_1) M^{(3)}(\mathbf{k},\mathbf{q}_2,-\mathbf{q}_2) P_L(q_1)P_L(q_2)\,, \\
    P_{A}^{(33b)}(k) &=& 6 \int_{\mathbf{q}_{12}} \left[ M^{(3)}(\mathbf{k}-\mathbf{q}_1-\mathbf{q}_2,\mathbf{q}_1,\mathbf{q}_2)\right]^2 P_L(|\mathbf{k}-\mathbf{q}_1-\mathbf{q}_2|)P_L(q_1)P_L(q_2) \label{eq:PA_33b}\, ,
\end{eqnarray} \normalsize
where $P_L$ is the linear power spectrum obtained using the \texttt{CAMB} code \cite{CAMB}. Each one of those one and two-loop terms are shown by dashed lines on the left and right panels of Fig.~\ref{fig:loopterms}, respectively. Notice that, on large-scales, the terms that are not proportional to the external leg (e.g. $P_A^{(22)}$, $P_A^{(24)}$ and $P_A^{(33b)}$) go to a constant, as a consequence of the $F$ independent term in the $M$ kernels.

The total one and two-loop contributions for the Taylor-$A$PT method can be seen by dashed lines on the top panel of Fig~\ref{fig:PT_z}. The three-loop contribution is shown in orange and it was calculated only for the first modes measured in the simulation suite\footnote{Calculating each point of the three-loop spectrum of $A$ for both the Taylor-$A$PT and EoM-$A$PT schemes is much more time consuming than for $\delta$ due to the symmetrizations needed by the new kernels. It is beyond this project's scope to analyse the full-shape spectrum at three loops, such that we only calculated it for the first modes. The first modes provide an overall order of magnitude of the three-loop contribution and indicate whether the theory is convergent on large scales.}. We compare the full PT contribution to the simulation measurements on the bottom left panel of Fig~\ref{fig:PT_z}.
As pointed out by  \cite{Wang:2011fj} and \cite{Neyrinck:2009fs}, the measured power spectrum for $A$ differs from the power spectrum for $\delta$ already on the large scales. This shift is normally referred to as a large-scale bias\footnote{One could in principle use the ratio between the variance of $A$ and $\delta$ to estimate this large-scale bias \cite{Neyrinck:2009fs, Repp:2016gxt}
\begin{eqnarray}
    b_A^2  \simeq \frac{\sigma_A^2}{\sigma_\delta^2} \,.
\end{eqnarray}
} and it depends on the smoothing scale $R$. 

Therefore the linear theory for $A$ already fails on the large scales and the loop diagrams start to contribute at low $k$. The largest scales are improved at one loop and our result agrees with \cite{Wang:2011fj}. However, one can see that even though the series terms get smaller, adding more loop-terms does not pull the theoretical prediction in the direction of the N-body simulation data. It somewhat resembles what is already observed for $\delta$ on intermediate scales \cite{Blas:2013aba,Konstandin:2019bay} and gives evidence that the series for $A$ is asymptotic already on the large scales. This motivates us to build in the following section a perturbative approach for $A$ that is not constructed on top of $\delta_R$.

\subsection{EoM-$A$PT}\label{sec:Apt}

Motivated by constructing a perturbative series for $A$ that is not built on top of $\delta$ and that has a better loop-convergence structure on the large scales, we proceed to an alternative approach. The idea is not to expand the log, but to construct a perturbative series for $A$ directly through the equations of motion. We label this approach as EoM-$A$PT. 

We start by writing the continuity and Euler's equations for a collisionless fluid, together with Poisson's equation:
\begin{eqnarray}
\partial_\tau \delta (\mathbf{x}, \tau)  + \nabla \cdot \left[ (1+\delta(\mathbf{x}, \tau) ) \mathbf{u} (\mathbf{x}, \tau) \right] &=& 0 \,, 
\\
\partial_\tau \mathbf{u}(\mathbf{x}, \tau)  + \mathcal{H}\mathbf{u}(\mathbf{x}, \tau)  + \mathbf{u}(\mathbf{x}, \tau) \cdot \nabla \mathbf{u}(\mathbf{x}, \tau)  &=& - \nabla \Phi(\mathbf{x}, \tau)  \,,
\\
\nabla^2 \Phi(\mathbf{x}, \tau)  &=& \frac{3}{2}\mathcal{H}^2\delta(\mathbf{x}, \tau)  \,,
\end{eqnarray}
where $\mathbf{u}(\mathbf{x}, \tau)$ is the comoving velocity of the fluid and $\Phi(\mathbf{x}, \tau)$ is the comoving gravitational potential.
After defining $\theta = \nabla  \mathbf{u}$, we take the divergence of Euler's equation, replace $\nabla^2 \Phi$ and convolve a filter $W$ in each equation\footnote{Smoothing the equations of motions is the typical procedure to construct the EFTofLSS \cite{Baumann:2010tm}. Its smoothing is typically done over the non-physical scale $\Lambda$ and the counter-terms cancel out this dependence on $\Lambda$. Differently, the smoothing here has been done over a physical scale $R$. Even though the counter-terms will appear to parametrize the small-scale physics, they do not need to cancel out the $R$ dependence. We discuss more the scales of the problem in Sec.~\ref{sec:eft}. }
\begin{eqnarray}
\partial_\tau \delta_R  + \theta_R   
+ W_R \star \left( \delta_R \theta_R + \frac{\nabla \theta_R}{\nabla^2}  \cdot \nabla \delta_R  \right) =  \textrm{c.t.}  \label{eq:Aeom_cont}\,, 
\\
\partial_\tau \theta_R  + \mathcal{H} \theta_R  + \frac{3}{2}\Omega_m\mathcal{H}^2\delta_R =
- W_R \star \left[ \nabla \cdot \left( \mathbf{u}\cdot \nabla \mathbf{u} \right) \right] + \textrm{c.t.} \label{eq:Aeom_euler}\,,
\end{eqnarray} 
where we have decomposed each field in its long and short-wavelength parts (as done in \cite{Carroll:2013oxa}). $\textrm{c.t.}$ states for terms with at least one short-wavelength component and that appear as counter-terms after being integrated out \cite{Carroll:2013oxa}.  Using Eq.~(\ref{eq:Adef}), we have
\begin{eqnarray}
 \partial_\tau A    
+ e^{-A} W_R \star \left[ e^A \left( \theta_R + \frac{\nabla \theta_R}{\nabla ^{2}}  \cdot  \nabla A \right) \right] =  \textrm{c.t.} \,, \label{eq:afiltering}\\
\partial_\tau \theta_R + \mathcal{H} \theta_R  
+ \frac{3}{2}\Omega_m\mathcal{H}^2(e^A-1) = 
- W_R \star \left[ \nabla \cdot \left( \mathbf{u}\cdot \nabla \mathbf{u} \right) \right] + \textrm{c.t.} \,. \label{eq:tfiltering}
\end{eqnarray}
 After dropping out the subscript $R$ in $\theta _{R}$, the equations of motion in Fourier space can be written as 
\begin{eqnarray} \label{eq:Asystem}
\partial_\tau A (\mathbf{k},\tau)+ \theta (\mathbf{k},\tau)   &=& 
\\
- W(kR)\int_{\mathbf{q}_{12}} \delta_{\rm D}\left(\mathbf{k} - \mathbf{q}_{12}\right)\ \theta_\mathbf{q_1}A_{\mathbf{q_2}} \alpha'(\mathbf{q_1},\mathbf{q_2})  &-& T_1[A, \theta](\mathbf{k},\tau) - T_2[A, \theta](\mathbf{k},\tau) + \textrm{c.t.}  \nonumber \,,
\\
\partial_\tau \theta (\mathbf{k},\tau) + \mathcal{H} \theta (\mathbf{k},\tau) + \frac{3}{2}\Omega_m\mathcal{H}^2 A (\mathbf{k},\tau) &=&  \label{eq:Thetasystem}
\\
- W(kR)\int_{\mathbf{q}_{12}} \delta_{\rm D}\left(\mathbf{k} - \mathbf{q}_{12}\right)\theta_{\mathbf{{q_1}}}\theta_{\mathbf{q_2}}\beta(\mathbf{q_1},\mathbf{q_2})
&-& T_3[A, \theta] (\mathbf{k},\tau)+ \textrm{c.t.}  \nonumber \,,
\end{eqnarray}
where the exponential in Eq.~(\ref{eq:tfiltering}) was expanded at all orders and
\begin{equation} \label{eq:newalpha}
  \alpha' (\mathbf{q_1},\mathbf{q_2}) =   \frac{(q_1^2+q_2^2)(\mathbf{q}_1\cdot\mathbf{q}_2)}{2q_1^2q_2^2} = \alpha (\mathbf{q_1},\mathbf{q_2}) - 1 \,,
\end{equation}
with $\alpha$ and $\beta$ being the usual PT symmetrized vertices for the $\delta$ field. Notice that the main difference compared to the usual PT procedure for $\delta$ is that the vertex $\alpha$ is changed and three new terms $T_1[A, \theta]$, $T_2[A, \theta]$ and $T_3[A, \theta]$ appear.  Also now $\alpha'(\mathbf{k},-\mathbf{k}) = -1 $, indicating a coupling between the modes even in the limit of $k \rightarrow 0$.
The temporal dependence of Eqs~(\ref{eq:Asystem}) and (\ref{eq:Thetasystem}) still drops out after using the EdS approximation. 

We comment now on each one of the $T_{n}$ terms. $T_1[A, \theta]$ and $T_2[A, \theta]$ represent the two contributions in the square brackets of Eq.~(\ref{eq:afiltering}) and they arise because we can not cancel out $e^{-A}$ and $e^{A}$ in Eq.~(\ref{eq:afiltering}) due to the filter convolution. In Fourier space, those terms are given by
\begin{eqnarray} 
\label{eq:T1}
T_1[A, \theta] (\mathbf{k}) &=&  \int_{\mathbf{q}_{123}} \delta_{\rm D}(\mathbf{k}-\mathbf{q}_{123})  \times 
\\ &&\sum_{n,m,n+m>0} \frac{\mathcal{C}[(-A)^n]_{\mathbf{q}_1}}{n!}   \,W( |\mathbf{k}-\mathbf{q}_1| R)\, 
 \frac{\mathcal{C}[A^m]_{\mathbf{q}_2}}{m!} \,\theta_{\mathbf{q}_3}  \,, \nonumber
\\ 
\label{eq:T2}
T_2[A, \theta] (\mathbf{k}) &=&   \int_{\mathbf{q}_{1234}} \delta_{\rm D}(\mathbf{k}-\mathbf{q}_{1234}) \times
\\ &&\sum_{n,m, n+m > 0} \frac{\mathcal{C}[(-A)^n]_{\mathbf{q}_1}}{n!}   \,W( |\mathbf{k}-\mathbf{q}_1| R)\,   
\frac{\mathcal{C}[A^m]_{\mathbf{q}_2}}{m!} \, \frac{\theta_{\mathbf{q}_3} }{q_3^2}\, \mathbf{q}_3 \cdot \mathbf{q}_4 \,A_{\mathbf{q}_4} \,. \nonumber
\end{eqnarray}
$\mathcal{C}[A^m]_{\mathbf{q}}$ denotes a convolution of $m$ fields $A$ evaluated with external momenta $\mathbf{q}$. Note that both $T_1$ and $T_2$ are zero in the limit $W \rightarrow 1$. $T_1$ already contributes at second order in perturbation theory, while $T_2$ starts to contribute only at third order. Finally, $T_3[A, \theta]$ comes from the non-linear terms in the exponential expansion in Eq.~(\ref{eq:tfiltering}) and it is calculated by
\begin{eqnarray}
\label{eq:T3}
  T_3[A, \theta] (\mathbf{k}) &=& \frac{3}{2}\Omega_m\mathcal{H}^2 \sum_{n=2}\frac{1}{n!}\left[\int_{\mathbf{q}_{1\dots n}}  \delta_{\rm D}\left(\mathbf{k} -\mathbf{q}_{1\dots n} \right) A_{\mathbf{q}_1}\dots A_{\mathbf{q}_n}\right]  \,.
\end{eqnarray}

Performing again the expansion~(\ref{eq:deltapt_pert}), we find modified versions of the kernels ${F}^{(n)}$ and ${G}^{(n)}$. We call them respectively $\mathcal{F}^{(n)}$ and $\mathcal{G}^{(n)}$, such that
\begin{eqnarray}
    A^{(n)}(\mathbf{k}) &=& \int_{\mathbf{q}_{1\dots n}} \delta_{\rm D}(\mathbf{q}_{1\dots n} - \mathbf{k})\mathcal{F}^{(n)}(\mathbf{q}_1, \dots, \mathbf{q}_n)A^{(1)}_{\mathbf{q}_1}\dots A^{(1)}_{\mathbf{q}_n}  \,,
    \\
    \theta^{(n)}(\mathbf{k}) &=& \int_{\mathbf{q}_{1\dots n}}\delta_{\rm D}(\mathbf{q}_{1\dots n} - \mathbf{k}) \mathcal{G}^{(n)}(\mathbf{q}_1, \dots, \mathbf{q}_n)A^{(1)}_{\mathbf{q}_1}\dots A^{(1)}_{\mathbf{q}_n}  \,,
\end{eqnarray}
and using the shorthand notation $\mathcal{F}^{(n)}_{1,\dots, n}$ and $\mathcal{G}^{(n)}_{1,\dots, n}$ for $\mathcal{F}^{(n)}(\mathbf{q}_1, \dots, \mathbf{q}_n)$ and $\mathcal{G}^{(n)}(\mathbf{q}_1, \dots, \mathbf{q}_n)$,  their recursive relations are given by
\begin{eqnarray}
    \mathcal{F}^{(n)}_{1,\dots, n} &=& W(q_{1\dots n}R)\sum_{m=1}^{n-1}\frac{\mathcal{G}^{(m)}_{1,\dots, m}}{(2n+3)(n-1)} \times 
    \\
   && \left[ (2n+1)\alpha'(\mathbf{q}_{1\dots m},\mathbf{q}_{m+1\dots n}) \mathcal{F}^{(n-m)}_{m+1,\dots, n} + 2\beta(\mathbf{q}_{1\dots m},\mathbf{q}_{m+1\dots n}) \mathcal{G}^{(n-m)}_{m+1\dots n} \right]   \nonumber
   \\
    &+& \frac{(2n+1)[T_1^{(n)}(\mathbf{q}_1, \dots, \mathbf{q}_n) +T_2^{(n)}(\mathbf{q}_1, \dots, \mathbf{q}_n)]   + 2T_3^{(n)}(\mathbf{q}_1, \dots, \mathbf{q}_n)}{(2n+3)(n-1)} \nonumber\,,
\\
    \mathcal{G}^{(n)}_{1,\dots, n} &=& W(q_{1\dots n}R)\sum_{m=1}^{n-1}\frac{\mathcal{G}^{(m)}_{1,\dots, m}}{(2n+3)(n-1)} \times
    \\
    &&\left[ 3\alpha'(\mathbf{q}_{1\dots m},\mathbf{q}_{m+1\dots n}) \mathcal{F}^{(n-m)}_{m+1,\dots, n} + 2n \beta(\mathbf{q}_{1\dots m},\mathbf{q}_{m+1\dots n}) \mathcal{G}^{(n-m)}_{m+1,\dots, n} \right]  \nonumber
    \\ 
    &+& \frac{3\,[T_1^{(n)}(\mathbf{q}_1, \dots, \mathbf{q}_n) +T_2^{(n)}(\mathbf{q}_1, \dots, \mathbf{q}_n)]   + 2nT_3^{(n)}(\mathbf{q}_1, \dots, \mathbf{q}_n)}{(2n+3)(n-1)} \nonumber\,.
\end{eqnarray}
At linear order, $A^{(1)}$ must satisfy the linear version of Eq.~(\ref{eq:Adef})
\begin{equation} \label{eq:linearA}
    A^{(1)} (\mathbf{k}) = W(kR) \,\delta^{(1)} (\mathbf{k})\,,
\end{equation}
such that 
\begin{equation} 
    \mathcal{F}^{(1)}  = \mathcal{G}^{(1)} = W(kR)\,,
\end{equation}
and $T_i^{(n)}$ states for the expansion of $T_1[A, \theta]$, \, $T_2[A, \theta]$ and $T_3[A, \theta]$ at order $n$
\begin{eqnarray}
T_i(\mathbf{k}) = \sum_n T^{(n)}_i(\mathbf{k})\,.
\end{eqnarray}
We emphasize that when expanding each $T[A, \theta]$ term at some order, one needs to take into account all combinations of exponents and field perturbations. E.g. when calculating $T_1[A, \theta]$ at third order through Eq.~(\ref{eq:T1}) there will be five possibilities of exponents $(n,m)$: $(1,0),\, (0,1),\, (2,0),\, (1,1)$ and $(0,2)$. Each one of those cases contains several perturbations of $-A^n,A^m$ and $\theta$ that are third order in perturbation theory. After that, one also needs to consider all symmetrizations of the momenta. This severely increases the computational time needed by the EoM-$A$PT method if compared to the Taylor-$A$PT.

As a matter of comparison between both methods, we provide the explicit form of the second-order kernel
\begin{eqnarray}
    \mathcal{F}^{(2)}(\mathbf{q_1},\mathbf{q_2}) &=& W(q_{12}R)\, \frac{\mathcal{G}^{(1)}}{7}\left[ 5\alpha' \mathcal{F}^{(1)} + 2\beta \mathcal{G}^{(1)} \right] + \frac{5\,(T_1^{(1)} + T_2^{(2)}) + 2\,T_3^{(2)}}{7} \,.
\end{eqnarray}
 Since
\begin{eqnarray}
T_1^{(2)}(\mathbf{q_1},\mathbf{q_2}) &=&  -\frac{1}{2}[W(q_1)+W(q_2)]\,W(q_1R)\,W(q_2R) + W(kR) \,,
\\
T_2^{(2)}(\mathbf{q_1},\mathbf{q_2}) &=& 0 \,,
\\
T_3^{(2)}(\mathbf{q_1},\mathbf{q_2}) &=& \frac{3}{4}\,W(q_1R)\,W(q_2R)\,.
\end{eqnarray}
the final form of $\mathcal{F}^{(2)}$ is
\begin{eqnarray}
    \mathcal{F}^{(2)}(\mathbf{q_1},\mathbf{q_2}) &=& F^{(2)}\,(\mathbf{q_1},\mathbf{q_2})\,W(q_{12}R)\,W(q_1)\,W(q_2) \\
    &-& \frac{1}{14}\,W(q_1)\,W(q_2) \,[\,5W(q_1)+5W(q_2) - 3]  \nonumber \,.
\end{eqnarray}
Notice that smoothing the equations of motion leads to a different result than the perturbative approach described in the Sec.~\ref{sec:deltapt}. The differences come mainly on where to insert the filter convolutions. In the limit $W\rightarrow 1$ we have
\begin{eqnarray}
    \mathcal{F}^{(2)}(\mathbf{q_1},\mathbf{q_2}) &=& F^{(2)} (\mathbf{q_1},\mathbf{q_2})- 1/2\,,
\end{eqnarray}
recovering Eq.~(\ref{eq:M2}) on the same limit. Performing the same calculation for $\mathcal{G}^{(2)}$ we find
\begin{eqnarray}
       \mathcal{G}^{(2)}(\mathbf{q_1},\mathbf{q_2}) &=& G^{(2)}(\mathbf{q_1},\mathbf{q_2}) \,.
\end{eqnarray}
 It is also straightforward to show that in this limit
\begin{eqnarray}
    \mathcal{F}^{(3)}(\mathbf{q_1},\mathbf{q_2},\mathbf{q_3}) &=&  F^{(3)} (\mathbf{q_1},\mathbf{q_2},\mathbf{q_3})- \frac{1}{3}[F^{(2)}(\mathbf{q_1},\mathbf{q_2})+2 \textrm{ sym}] + \frac{1}{3}   \,,
\end{eqnarray}
also recovering Eq.~(\ref{eq:M3}). The equivalence between both approaches in the limit $W\rightarrow 1$ is linked to two facts. First, the exponentials in Eq.~(\ref{eq:afiltering}) cancel out at each order, leading to $T_1 \rightarrow 0$ and $T_2 \rightarrow 0$. Second, $T_3$ recovers the Taylor expansion of the Taylor-$A$PT approach.

\begin{figure}[ht]
\centering
  \includegraphics[width=0.49\textwidth]{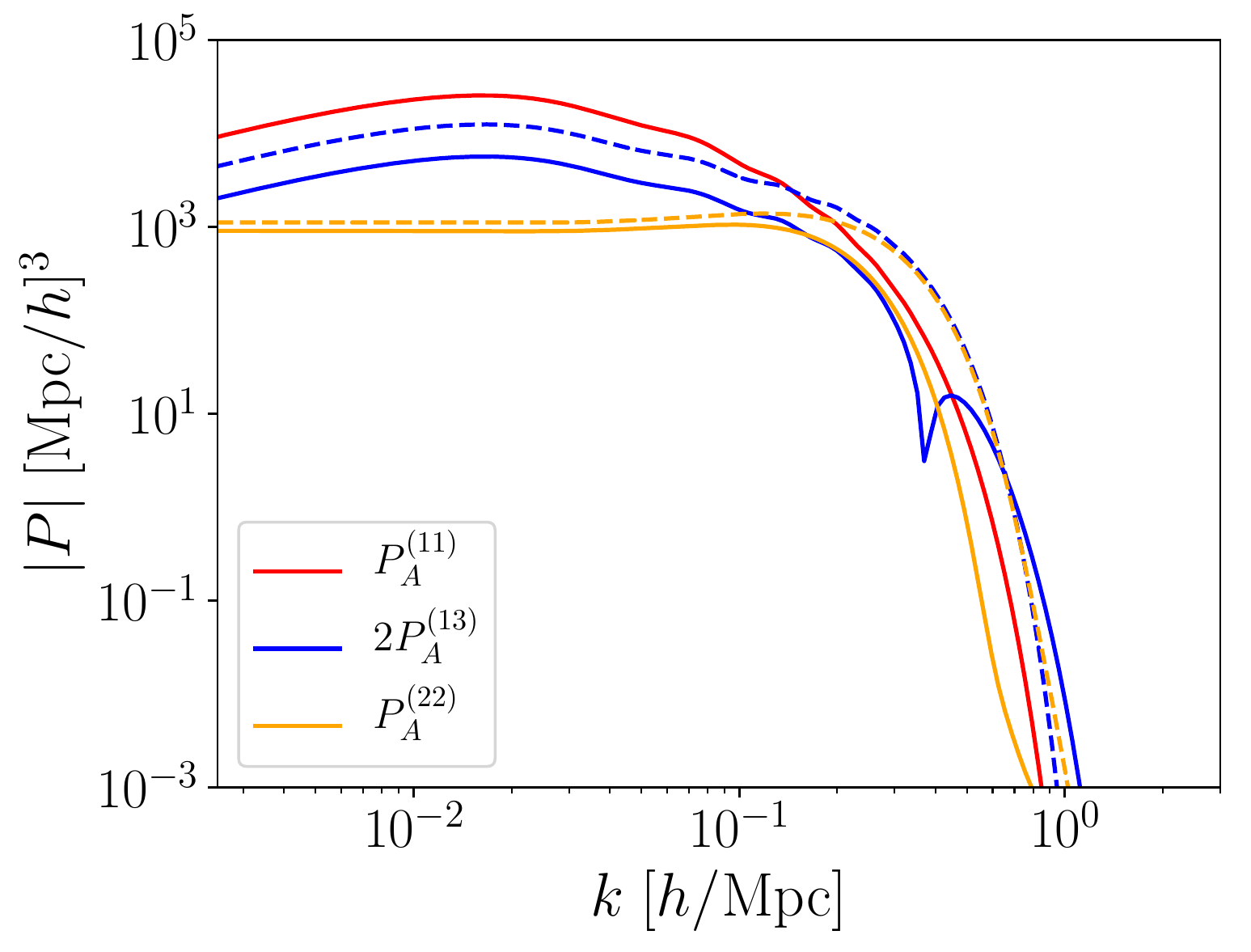}
  \includegraphics[width=0.49\textwidth]{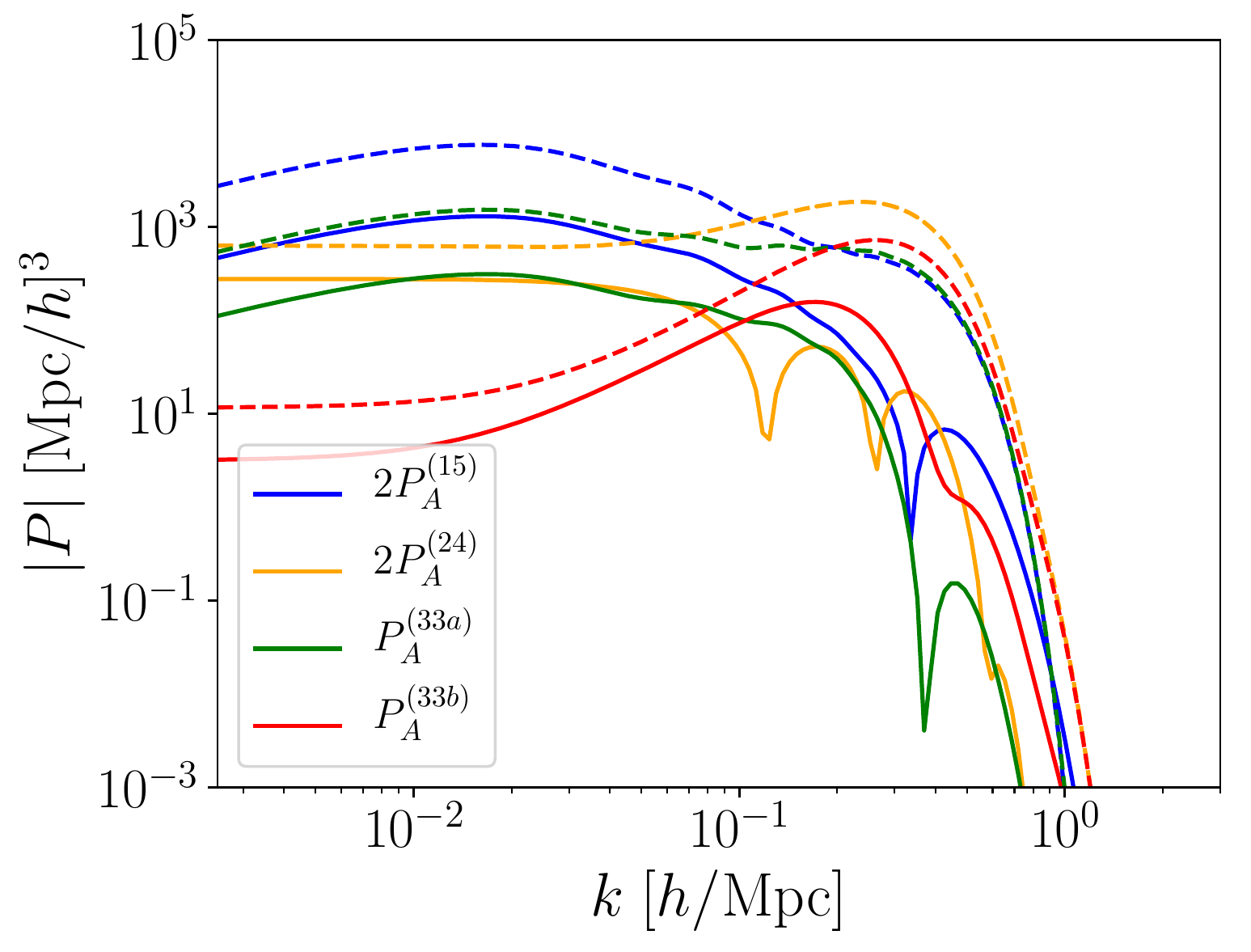}
\caption{\label{fig:loopterms}%
\small  Each loop term contributions for $A$ at $z=0$. On the left, the linear and one-loop terms for the Taylor-$A$PT scheme (dashed) described in Sec.~\ref{sec:deltapt} and for the EoM-$A$PT scheme (continuous) described in Sec.~\ref{sec:Apt}. On the right, the same but for the two-loop terms. The linear power spectrum is the same for both models.}
\end{figure}

The one and two-loops terms are calculated in the same way as Eqs.~(\ref{eq:PA_11}) -  (\ref{eq:PA_33b}), with the replacement $M\rightarrow\mathcal{F}$. We display each one of the one and two-loops terms calculated through the EoM-$A$PT method by solid lines in Fig.~\ref{fig:loopterms}. On large scales, both approaches described here and in Sec.~\ref{sec:deltapt} only differ by a constant shift\footnote{As will be discussed in the following sections, the low-$k$ limit of the one-loop theory is a constant that will be absorbed by free parameters.}. The shape of the curves for the EoM-$A$PT method and the Taylor-$A$PT method start to differ on intermediate $k$. 

\begin{figure}[ht]
\centering
  \includegraphics[width=0.49\textwidth]{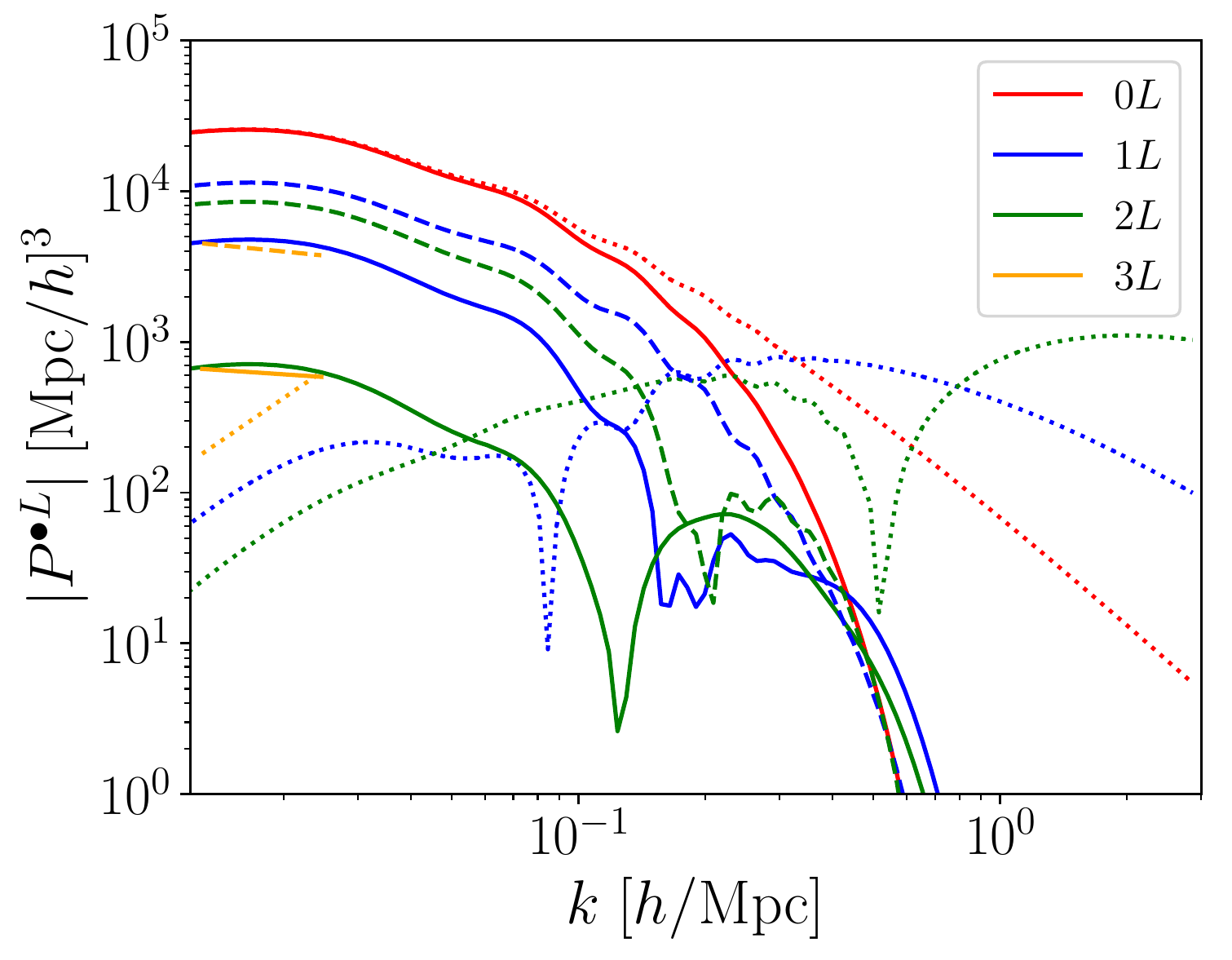}\\
  \includegraphics[width=0.49\textwidth]{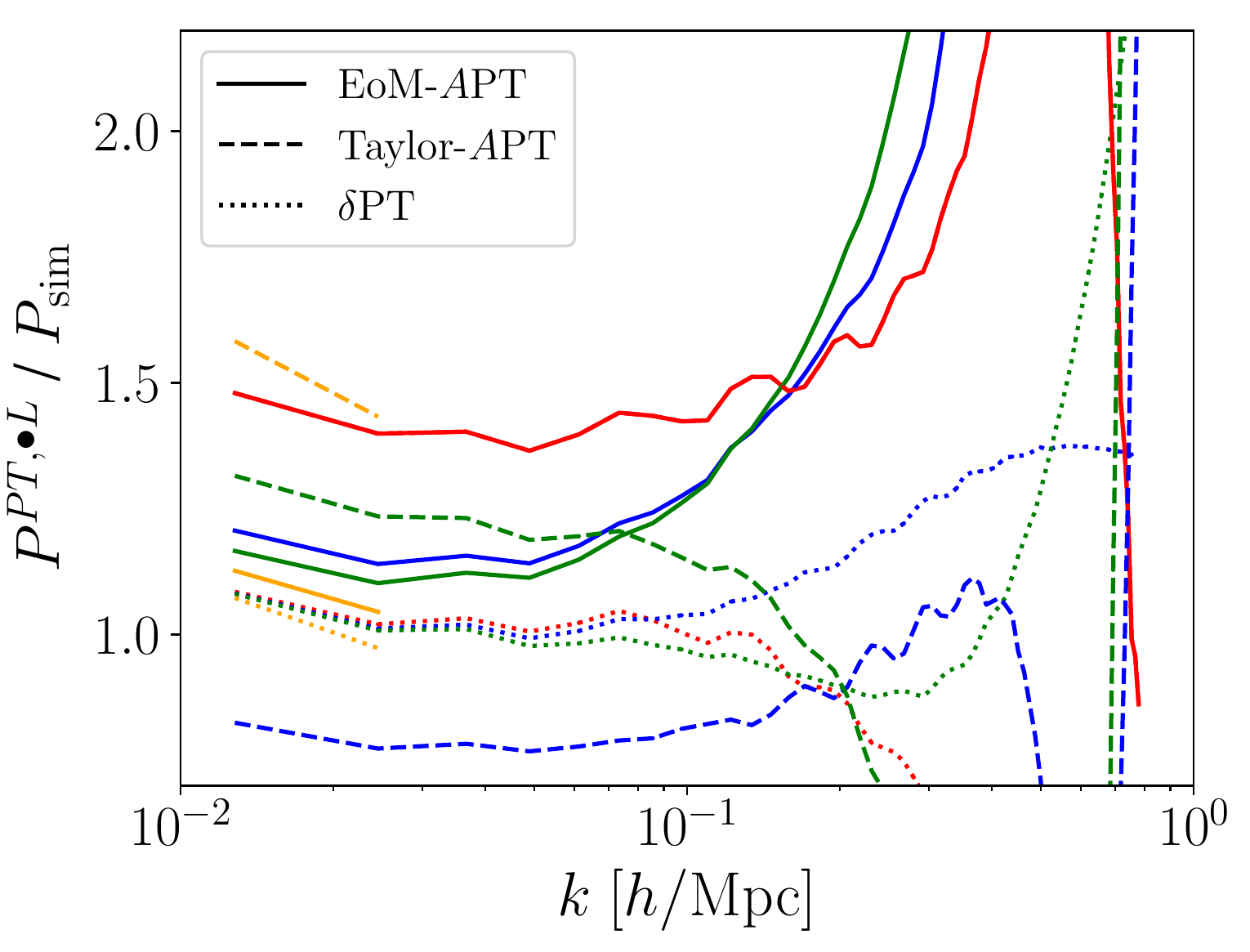}  
  \includegraphics[width=0.49\textwidth]{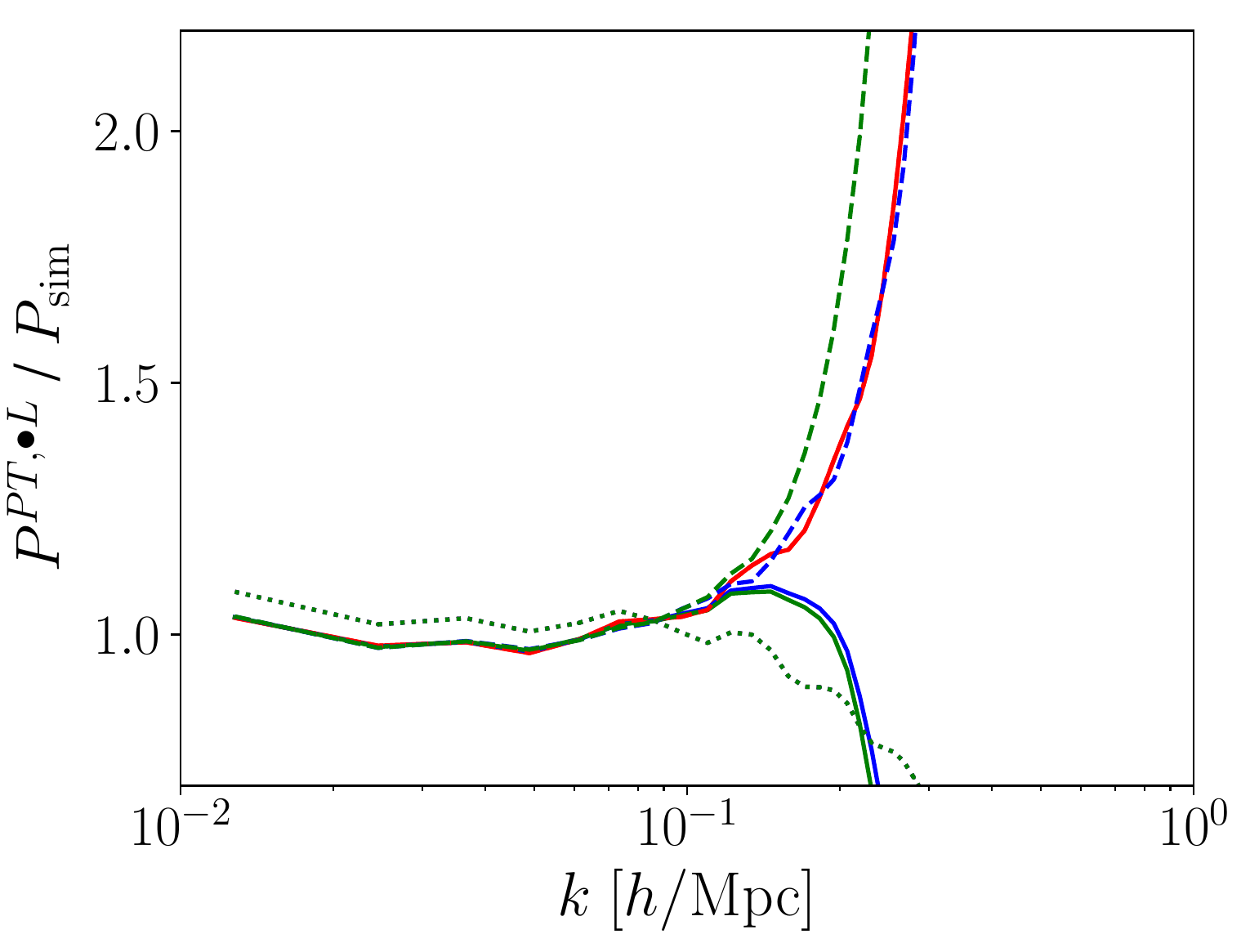}
\caption{\label{fig:PT_z}%
\small On the lop, the linear, one and two-loop contributions to the total PT power spectrum at $z=0$. We compare two different schemes: Taylor-$A$PT (dashed lines) and EoM-$A$PT method (solid lines). The PT prediction for $\delta$ is shown in dotted lines. The three-loop contribution for the first modes measured in the simulation are shown by orange lines. On the bottom left, we compare the total PT result to the numerical simulations.  On the bottom right panel, we show the same lines as in the left panel but after fixing the largest scales for the $A$ models. For that, we added the two free parameters pointed out in Sec.~\ref{sec:largescales}: a constant term (shot noise like) and a term proportional to $P_L$ (bias like). Notice that the solid and dashed lines are normalized by the power spectrum for $A$ while the dotted lines are normalized by the power spectrum of $\delta$.
}
\end{figure}

In Fig.~\ref{fig:PT_z}, we display the one, two and three-loop results using the EoM-$A$PT (continuous), the Taylor-$A$PT method (dashed) and the usual PT for $\delta$ (dotted). On the top panel we show each loop contribution to the total power spectrum at $z=0$, and on the bottom left panel we compare the result to simulations. Notice that the hierarchy of the EoM-$A$PT method follows $P^{1L}_A \gg P^{2L}_A \sim P^{3L}_A$, which indicates that the series converges better than the Taylor-$A$PT method. The EoM-$A$PT method also coherently approaches the simulation result on large scales. Moreover, the Taylor-$A$PT contributions change sign and its total three-loop result departs from the N-body simulation measurement. It strongly indicates that the PT series within the Taylor-$A$PT scheme is asymptotic even on large scales, while using the EoM-$A$PT scheme ameliorate the convergence at $z=0$. In the next section, we comment on how to fix the large scales. Furthermore, Fig.~\ref{fig:PT_z}  highlights another important point in the PT for $A$: it is not perturbative in $k$. It can be seen by the shape of the loop corrections: whilst the loop corrections for $\delta$ are smaller on larger scales it is not the case for $A$. This makes the tree level prediction for the field $A$ to be not accurate even for small values of $k$.

\subsection{Fixing the large-scales and the propagator} \label{sec:largescales}

\begin{figure}[ht]
\centering
  \includegraphics[width=0.49\textwidth]{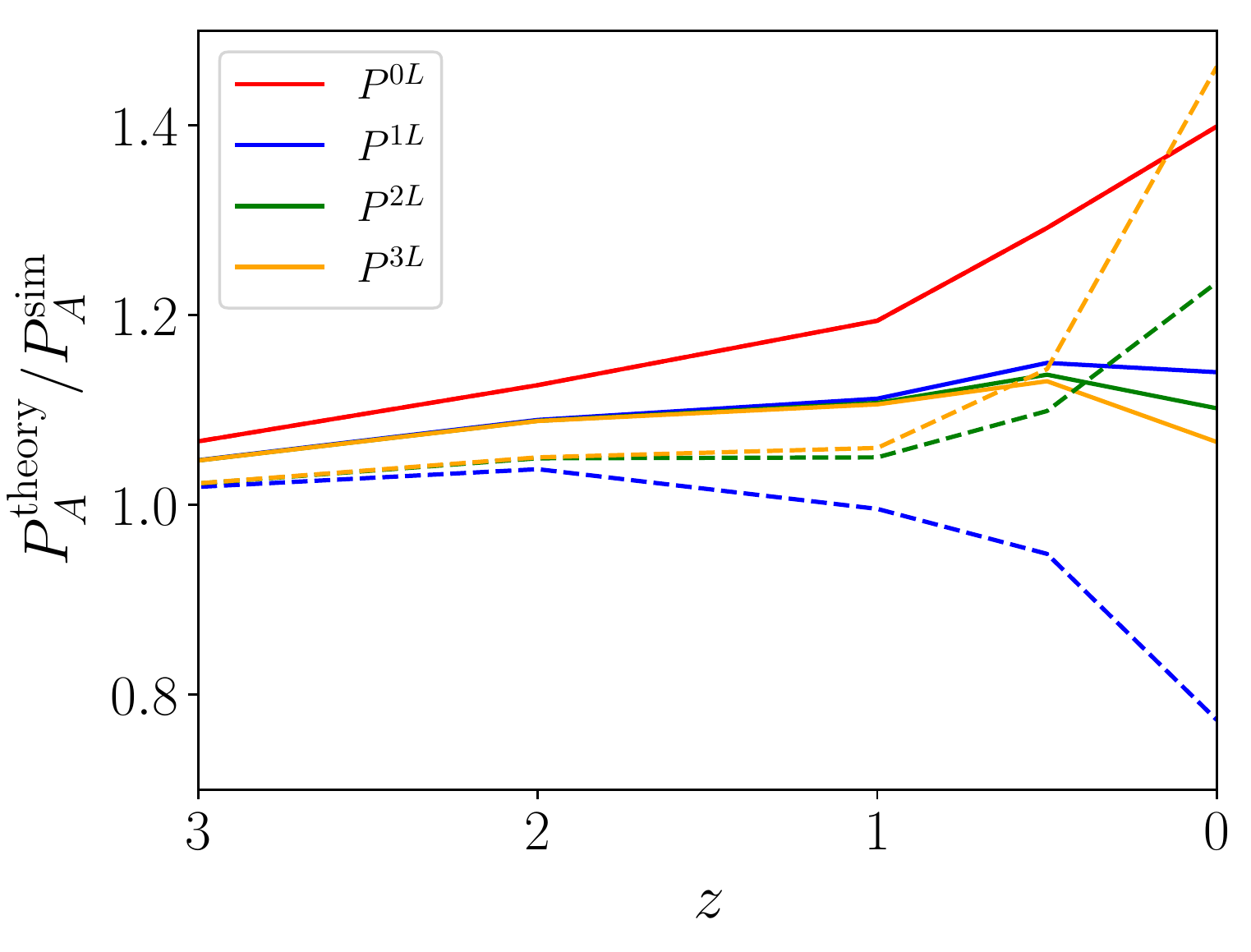}
\caption{\label{fig:bias_largescale}%
\small The ratio between the large-scale prediction for $A$ and the simulation result as a function of the redshift. In solid lines the theoretical prediction was calculated using EoM-$A$PT and in dashed lines we used the Taylor-$A$PT scheme. The cosmic variance for this mode ($k = 0.024 $ Mpc$^{-1}$h) is of $1.3\%$.}
\end{figure}

We now focus on understanding the low-$k$ behaviour of the power spectrum for $A$ and how it can be fixed.   
In Fig.~\ref{fig:bias_largescale} we compare the theoretical power spectrum on large scales ($k = 0.024 $ Mpc$^{-1}$h) to simulations for different redshifts. 
We compare the two different schemes presented in this work:  
EoM-$A$PT (solid lines) and Taylor-$A$PT method (dashed). At $z=3$ the one-loop PT contribution is the most relevant and both methods correctly predict the power spectrum with $\sim5\%$ error. 
The Taylor-$A$PT method performs better at high $z$. 
At low redshift, however, the Taylor-$A$PT method severely departs from the N-body result. 
Notice also that its loop-contributions change sign and that the three-loop prediction blows up at $z=0$. This series is organized on top of $\delta$ perturbations and carries the same convergence issues of the usual perturbation theory for $\delta$ \cite{Blas:2013aba} but already on the largest scales.

The large-scale prediction using the EoM-$A$PT method is more consistent across all redshifts. Even though the series error is larger on high $z$, its loop-structure seems to resemble what is expected from a well-behaved perturbative series: each loop contribution getting smaller and not changing sign. 

Two points raised by Fig.~\ref{fig:bias_largescale} need to be clarified: first, the reason why the perturbative approach fails at $z=0$ for the Taylor-$A$PT scheme and behaves better in the EoM-$A$PT scheme and second, how we can fix the large scales to achieve better precision for both schemes. 

Within the Taylor-$A$PT scheme, the contribution of $R$ to the large scales can be seen already in the shape of $M^{(2)}$ and $M^{(3)}$, Eqs.~(\ref{eq:M2}) and (\ref{eq:M3}). The last term in both cases is an $F$ independent term that leads to a relevant contribution in the loop-integrals and that is suppressed when the internal momentum is larger than $R^{-1}$.\footnote{The sensitivity of the largest modes of $A$ on small scales perturbations have been numerically shown in \cite{Neyrinck:2013iza}. } As already pointed out by \cite{Wang:2011fj} and recently by \cite{Philcox:2020fqx} for the marked transformations, the $F$ independent term in the kernels provides an important correction on small $k$. $P_{A}^{(22)}$ contributes to the large scales with a constant term that works as a shot-noise and $P_{A}^{(13)}$ adds a term proportional to $P(k)$ that behaves as a linear bias\footnote{See Appendix~\ref{app:scaling} for an explicit calculation.}. 

\begin{figure}[ht]
\centering
  \includegraphics[width=0.49\textwidth]{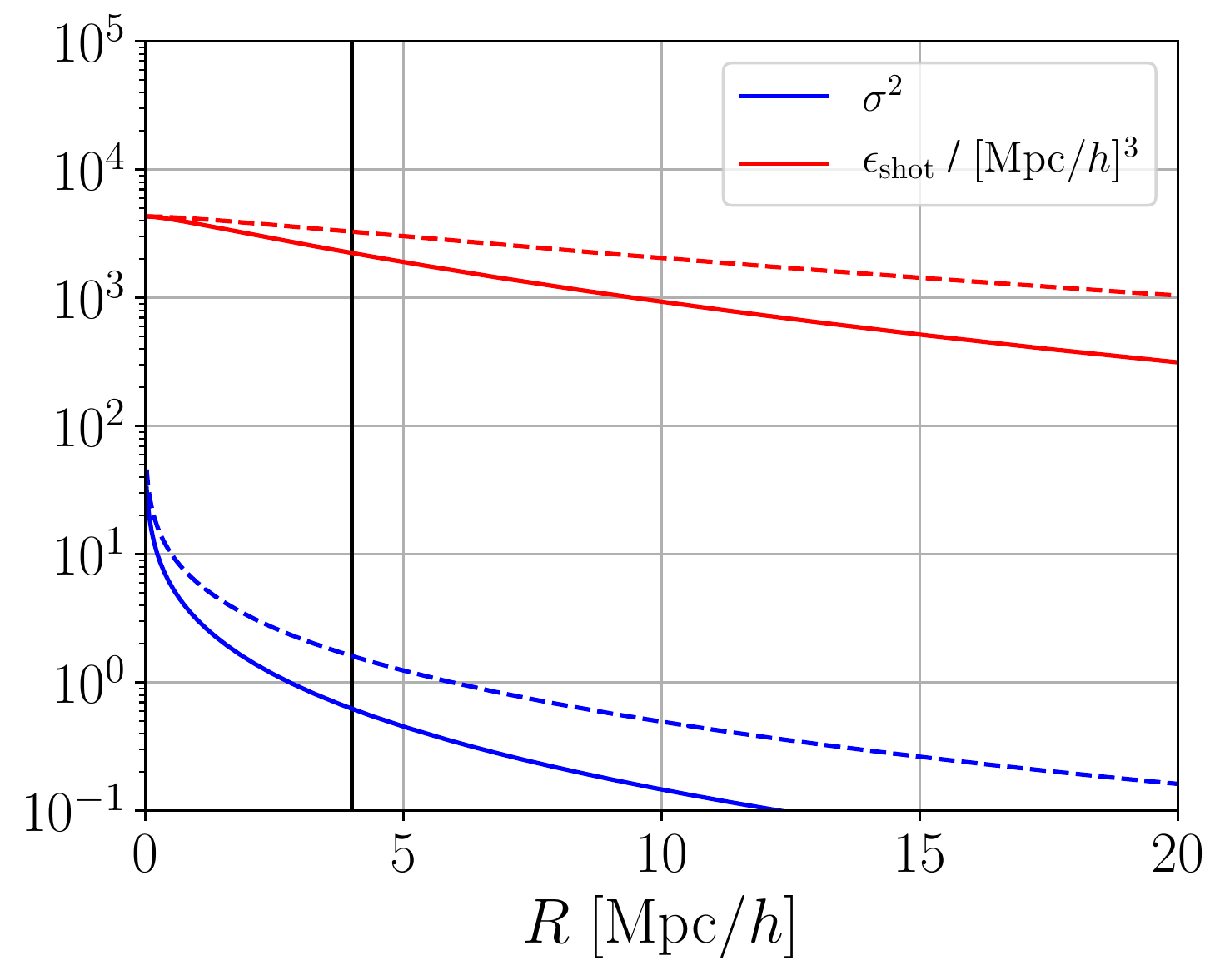}
\caption{\label{fig:variance}%
\small Parameters that control the size of the loop integrals for the log field on large scales (see Eqs.~(\ref{eq:eps}) and (\ref{eq:variance})). The parameters were calculated using a Gaussian filter (solid lines) and a top-hat in position space filter (dashed). The black vertical line shows $R = 4 $ Mpc h$^{-1}$, value adopted in this work.}
\end{figure}

If perturbation theory works well, one would expect that summing up all loop contributions for the series would reproduce the correct large-scale bias and propagator.
We see by Eqs.~(\ref{eq:P22}) and (\ref{eq:P13}) that the $P_{A}^{(22)}$ and $P_{A}^{(13)}$ contributions on large scales are respectively proportional to
\begin{eqnarray} \label{eq:eps}
   \epsilon_{\rm shot}(R) &\equiv& \int_\mathbf{q}  \left[P_L(q)\right]^2W^4(qR)  = 2\lim_{k\to0} P_{A}^{(22)}(k) \,, \\ \label{eq:variance}
   \sigma^2 (R) &\equiv& \int_\mathbf{q}  P_L(q)\,W^2(qR)  = \lim_{k\to0} \frac{P_{A}^{(13)}(k)}{P_L(k)}   \,.
\end{eqnarray}
 We show $\sigma^2$ and $\epsilon_{\rm shot}$ in Fig.~\ref{fig:variance} for the Gaussian filter (solid) and for the for a top-hat in position space filter (dashed). 
For $R = 4$ Mpc h$^{-1}$ and a Gaussian filter, which were adopted in this work, we have 
 \begin{eqnarray}
     \epsilon_{\rm shot}(R=4) = 2230 \,\,\, (\textrm{Mpc/h})^3  \,, \quad \textrm{and} \quad \sigma^2 (R=4) = 0.62 \lesssim 1     \,,
 \end{eqnarray}
 which agree with the values of $P_{A}^{(22)}$ and $P_{A}^{(13)}$ shown in the left panel of Fig.~\ref{fig:loopterms}. 

At $n$-loop order, the $F$ independent part of the kernels makes the terms proportional to an external leg $P^{(1,2n+1)}_A$ to scale approximately as $P(k)(\sigma^2)^n$ and the terms with no external leg to be roughly proportional to $\epsilon_{\rm shot}(\sigma^2)^{n-1}$.  Since $\sigma^2<1$, the perturbative series is therefore expected to be convergent
for a Gaussian filter with $R = 4 $ Mpc h$^{-1}$. It is indeed observed in the top panel of Fig.~\ref{fig:PT_z}  that each loop term for the Taylor-$A$PT approach is getting smaller, even though this is not a sufficient argument to guarantee the convergence of the series. Increasing $R$ would speed up the convergence but suppress the signal on larger scales and using a top-hat filter would increase $\epsilon_{\rm shot}$ and $\sigma^2$ and deteriorate the series's convergence.

Notice however that even though the Taylor-$A$PT terms get smaller, they do not seem to be converging to the simulation result. By the bottom left panel of Fig.~\ref{fig:PT_z}, one can see that adding the three-loop term induced the series away from the expected N-body result. This result is not surprising in face that the series for $\delta$ is not convergent on intermediate scales \cite{Blas:2013aba, Konstandin:2019bay} and those scales ($k < R^{-1}$) lead to an important contribution to the large scales of $A$. The asymptotic behaviour of the $\delta$ series is transposed to the Taylor-$A$PT series. 
 
The convergence results for the EoM-$A$PT method are indeed better at low $z$. It is possible to see by the bottom left panel of Fig.~\ref{fig:PT_z} that the EoM-$A$PT scheme converges faster. It is a consequence of the fact that all internal momenta are suppressed by a filter (see the linear spectrum for $A$ in Eq.~(\ref{eq:linearA})). Moreover, the extra convolution in the velocity source in Eq.~(\ref{eq:afiltering}) generates the $T_1[A, \theta]$ and $T_2[A, \theta]$ terms that lead to relevant corrections on the scaling of the loops. Even though it seems that smoothing directly through the equations of motion is in better agreement with the simulation data  at low $z$, there is no guarantee that adding infinite loop terms would lead to the correct amplitude on large scales. Moreover, the fact that the Taylor-$A$PT works better than EoM-$A$PT for high $z$ suggests that the extra smoothing of the EoM-$A$PT in that redshift is too aggressive. It motivates us to look for a systematic method to fix both Taylor-$A$PT and EoM-$A$PT schemes on large scales at all redshifts.   

On the bottom right panel of Fig.~\ref{fig:PT_z} we compare the theory with simulated data after adding two free parameters
    \begin{equation} \label{eq:free}
        \{ k^0, \,P_L(k)\} \,.
    \end{equation}
One, a constant term $P_{\rm shot}$ (proportional to $k^0$) that fixes diagrams like $P_{A}^{(22)}$; other, a term like $b_1^2 P_L(k)$ that fixes diagrams as $P_{A}^{(13)}$, which are corrections to the propagator. After adding those terms, both theories for $A$ reach similar non-linear scales as the perturbation theory for $\delta$ (dotted lines). We explicitly checked that this result is also reproduced at high $z$.

Notice that both the EoM-$A$PT and the Taylor-$A$PT models work equally good at all loop-orders after adding those free terms.  As pointed out by \cite{Philcox:2020srd} for the marked field, the inclusion of those terms can embrace a resummation of all loop-order terms\footnote{For the marked field the series is expected to converge faster due to the expansion coefficients. But there is no indication whether it converges to the value measured by N-body simulations and a more in-depth investigation is required.}. Here we can see that those terms are not only doing this resummation job but they also parametrize the correct scaling and fix an inherent problem of the perturbative theory for $A$, namely that it is asymptotic and fails already on the largest scales for $A$. 
    
We now investigate the loop corrections to the linear propagator akin it was done for $\delta$ in \cite{Crocce:2005xy,Crocce:2005xz} and in \cite{Wang:2011fj}, where they calculated the resummed propagator contributions for $A$ at one loop. Their scheme constructs a non-linear propagator by resumming all diagrams proportional to one external leg. The non-linear propagator $G$ is defined according to
\begin{eqnarray}
     \Bigg \langle \frac{\partial \phi_{\rm nl} (\textbf{k})}{\partial \phi_{\rm lin} (\textbf{k}')} \Bigg \rangle  = (2\pi)^3 \delta_{\rm D}(\mathbf{k} - \mathbf{k}') \,G(k) \,,
\end{eqnarray}
in which $\phi = {A,\delta}$. This propagator can be numerically calculated as a correlation between the initial conditions $\phi_{\rm lin}$ and the late time field $\phi_{\rm nl}$ \cite{Crocce:2005xz}. Theoretically, the corrections to the propagator are given by the PT terms coming from a contraction with an external leg. At one, two and three loops it is given by
\begin{eqnarray}
    G^{1L}(k) &=& \frac{P_L(k) + P^{(13)}(k)}{P_L(k)} \,,
    \\
    G^{2L}(k) &=& \frac{P_L(k) + P^{(13)}(k) + P^{(15)}(k)}{P_L(k)}\,,
        \\
    G^{3L}(k) &=& \frac{P_L(k) + P^{(13)}(k) + P^{(15)}(k) + P^{(17)}(k)}{P_L(k)}\,.
\end{eqnarray}

\begin{figure}[ht]
\centering
  \includegraphics[width=0.49\textwidth]{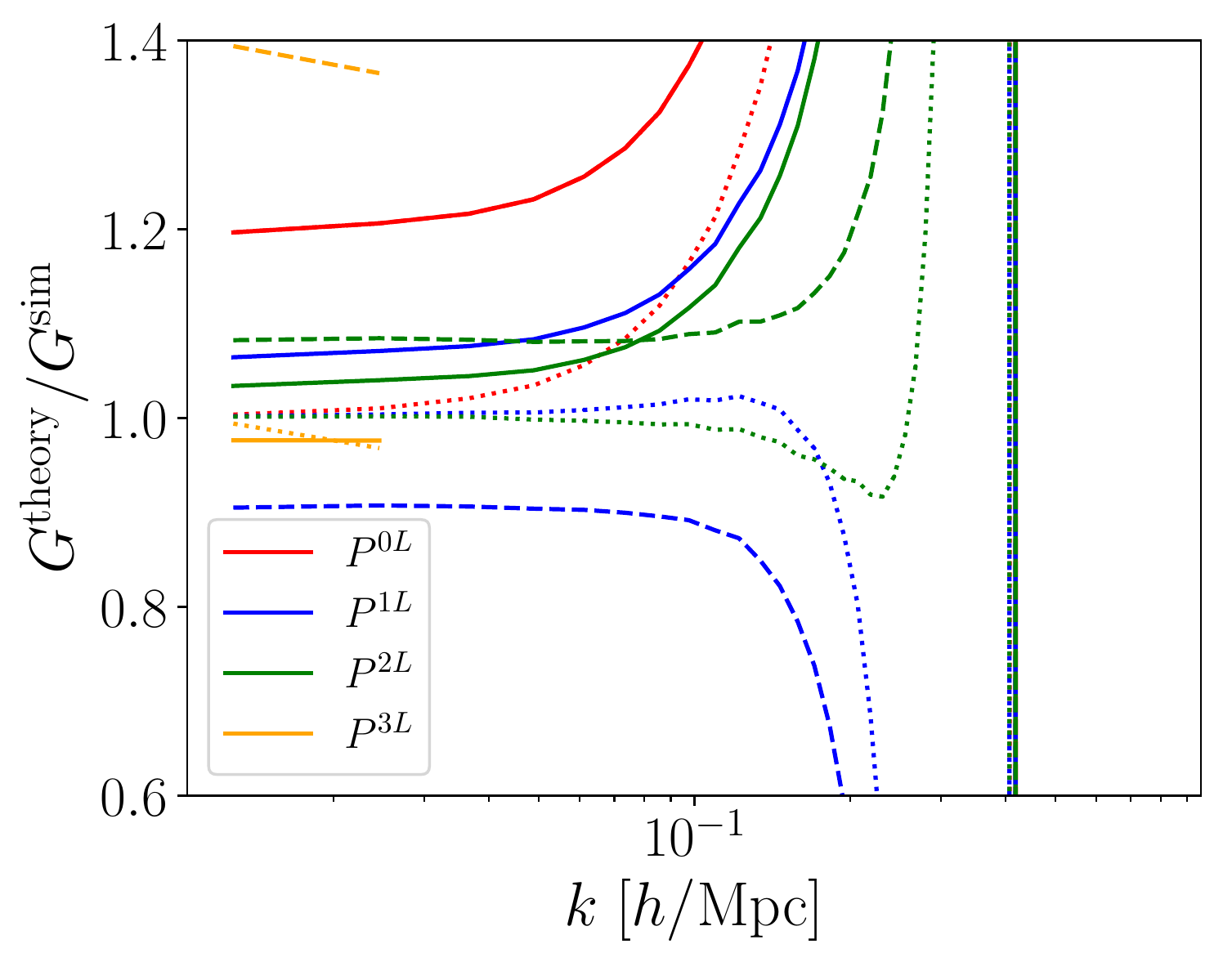}
  \includegraphics[width=0.49\textwidth]{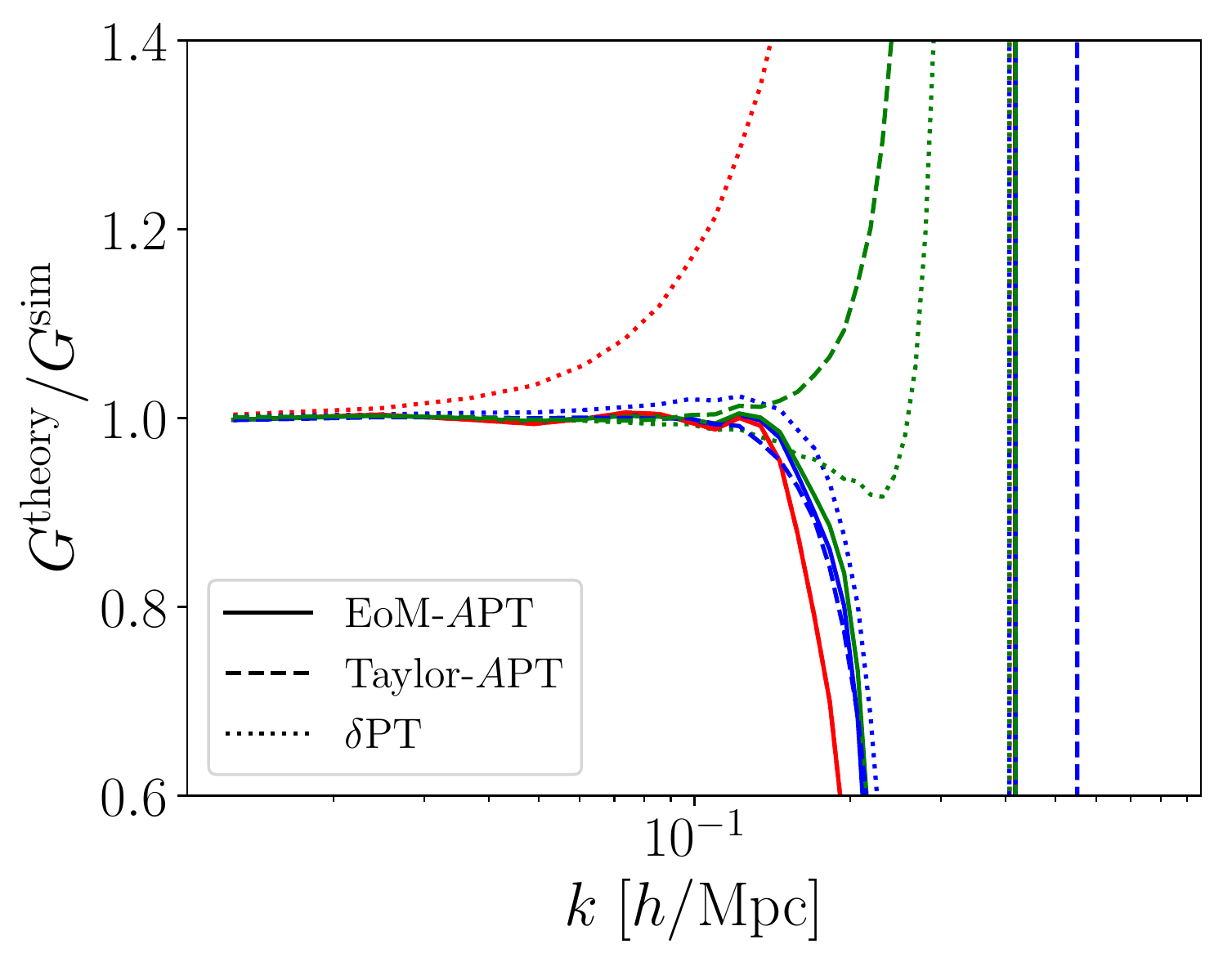}
\caption{\label{fig:propagator}%
\small On the left, the ratio between the theoretical prediction for the propagator and the propagator measured in N-body simulations at $z = 0$.  We show the result for three theories:  EoM-$A$PT (solid), Taylor-$A$PT (dashed) and for $\delta$ (dotted). On the right, the same but the predictions for the $A$'s power spectrum are corrected by a bias and a shot noise term fitted on large scales ($k<0.1 $ Mpc$^{-1}$h). Notice that the solid and dashed lines are normalized by the power spectrum for $A$ while the dotted lines are normalized by the power spectrum of $\delta$. }
\end{figure}

On the left panel of Fig~\ref{fig:propagator} we show the ratio between the theoretical prediction and the simulated results for $G$ calculated according to \cite{Crocce:2005xz} at $z=0$. Notice that the propagator resembles the non-convergent behaviour of the Taylor-$A$PT perturbative series: adding more loop terms is not leading the series to converge to the measured propagator. The EoM-$A$PT scheme reproduces the propagator corrections with better precision at low redshift and each loop-correction monotonically improves the propagator prediction. We explicitly checked that at higher $z$, in contrast, the Taylor method performs better, resembling to what is seen for the full spectrum in Fig.~\ref{fig:bias_largescale}.    

On the right panel of Fig~\ref{fig:propagator} we display the one and two-loop calculations when fitting the two free-parameters in Eq.~(\ref{eq:free}) on scales $k<0.1 $ Mpc$^{-1}$h.
We can see that the inclusion of those terms leads to an equally good result on large scales both for the EoM-$A$PT and the Taylor-$A$PT schemes. The addition of those free parameters led to the same precision of the prediction for $\delta$ (dotted line) at one and two loops and it also reaches comparable scales.

Another relevant term that gives a sub-leading correction on large scales is a second-order shot noise contribution that scales as $k^2$. This term comes from a large-scale contribution of  $P_{A}^{(22)}$ (see Appendix~\ref{app:scaling}) and we explicitly checked that its inclusion substantially improves the $A$ spectrum on intermediate $k$. We therefore include this free-parameter in the theory.

There is an important difference between the terms
    \begin{equation} \label{eq:free2}
        \{ k^0, k^2, P_L\} \,,
    \end{equation}
and the EFT counter-terms that fix the UV dependence. The former parametrize the dependence on scales $<R^{-1}$ and fix the offset of the theory and the N-body simulation result on large scales. They do not add a cutoff dependence since those are suppressed at $R^{-1}$ (see Appendix \ref{app:scaling}). In contrast, the counter-terms absorb the UV physics and cancel out the $\Lambda$ dependence. They are the topic of the following section.

\subsection{Effective field theory} \label{sec:eft}
We now proceed to calculate the effective field theory corrections for the logarithm field $A$.
Through a set of counter-terms, the effective field theory set up parametrizes how the small-scale physics affect other scales and at the same time it fixes the cutoff dependence of the PT integrals. Before calculating the UV dependence of each term, we discuss the four different scales that appear in the EFT framework and the hierarchy between them. The four scales are
\begin{enumerate}
    \item $k_\ast$, the scales we want the EFT to be predictive for;
    \item $R^{-1}$, the physical smoothing scale that defines $A$ in Eq~(\ref{eq:Adef});
    \item $k_{\rm NL}$, the scale where the perturbative approach breaks down;
    \item $\Lambda$, the non-physical cutoff used in the theory that appears as the upper limit of the integrals in Eqs.~(\ref{eq:PA_11})-(\ref{eq:PA_33b}).
\end{enumerate}

When calculating the EFT for $\delta$, the contribution of the smoothing scale $R$ used for the density grid in the N-body simulation is exponentially suppressed on scales $k_\ast$, such that it has in practice no effect on the problem. The only demand is therefore that
\begin{equation}
    k_\ast < k_{\rm NL} \quad \textrm{and } \quad k_\ast < \Lambda \,.
\end{equation}
The most natural choice for $\Lambda$ would be $\Lambda \sim k_{\rm NL}$ \cite{Carroll:2013oxa}. However, since we do not know $k_{\rm NL}$ \textit{a priori}, what is typically done is to take a very high value for $\Lambda$ and checking whether this arbitrary choice is canceled out by the counter-terms. 

When considering the log fields, the scale $R$ will affect all the scales of the problem. Different of $\Lambda$, $R$ is physical and should not be integrated out. Taking $k_{\rm NL}$ to be roughly estimated by the scale on which the variance of $\delta$ gets larger than unit (see Fig.~\ref{fig:variance}), we need to have 
\begin{equation} \label{eq:nltoR}
    k_{\rm NL} > R^{-1} \,,
\end{equation}
in order to guarantee that the Taylor series does not diverge \cite{Philcox:2020srd}. 
In case the condition~(\ref{eq:nltoR}) is not satisfied, the free parameters that parametrize scales $< R^{-1}$ and the counter-terms that parametrize UV scales $> k_{\rm NL}$ will contain mixed information and some extra degeneracies might arise. 
Moreover, one might also require the condition
\begin{equation}
    k_\ast < R^{-1}\,
\end{equation}
aiming to guarantee that the filtering does not entirely wash out relevant information. 

We now discuss how to parametrize the UV contributions in the effective field theory framework.
The most general set of counter-terms is the one allowed by the equations of motion symmetries (contracted derivatives of the gravitational and velocity potential that still preserve Lorentz invariance). A reduced set of counter-terms can be constructed by checking the UV dependence of each one of the loop integrals and identifying which counter-terms are necessary to absorb the cutoff contribution \cite{Baldauf:2015aha}. 
 In Appendix~\ref{app:scaling} we carry out a careful discussion concerning the scaling of the kernels for the effective field theory for $A$. We find that at one-loop level, the counter-term that cancels out the UV dependence is the same as for $\delta$:
\begin{eqnarray}\label{eq:ct1}
        P_{A}^{c.t,1L}(k) &=& -c_s^2\left(k R\right)^{2}P_L(k) \,,
\end{eqnarray}
which when fitted together with the large scales corrections~(\ref{eq:free2}) 
\begin{eqnarray} \label{eq:plarge}
        P_{A}^{\textrm{large } \textrm{scales}} &=& \{ k^0,\, k^2,\, P_L\} \,,
\end{eqnarray}
leads to a set of four free parameters. It is vital to notice that the inclusion of those free parameters is already needed in order to connect the theory to observations in the form of a linear bias and a shot noise term. In that sense the inclusion of those terms does not add any new degrees of freedom.
Therefore, the final expression for the one-loop power spectrum both for the EoM-$A$PT and the Taylor-$A$PT schemes is
\begin{eqnarray} \label{eq:eft1L}
        P_{A}^{EFT, 1L}  &=& P_{A}^{0L} + P_{A}^{1L}  + P_{A}^{c.t,1L} + P_{A}^{\textrm{large } \textrm{scales}} \,, \\
        &=& \left[b_{1}^{2} - c_s^2\left(k R \right) ^2\right] P_{A}^{0L} + P_{A}^{1L} + c_0\,k^0 - c_{1}\left(k R\right)^2 \,.
\end{eqnarray}

At two-loops, the number of free parameters needed to fix the large scales grows to six due to the new terms that appear coming from the $F$ independent terms in the $M$ kernels (see Appendix~\ref{app:scaling}). They are:
     \begin{equation} 
        \{ k^0,\, k^2,\, k^4,\, P_L,\, P^{1L}_\delta,\, P_{\rm quad}\} \,.
    \end{equation}
with $P_{\rm quad}$ defined in Eq.~(\ref{eq:pquad}). The counter-terms needed for $A$ are
\begin{eqnarray}
    \{ k^2P_L, k^2P^{1L}_\delta, k^4,k^4P_L, k^2 P_{\rm quad},P_{\rm bispec} \} \,.
\end{eqnarray}
 They are the same as the counter-terms considered for $\delta$  (see Appendix~\ref{app:scaling}) with one extra term
\begin{gather} \label{eq:pbispec}
  P_{\rm bispec} (k) =  P_L(k)\int_q  P_L(q)F^{(2)}(\mathbf{k},\mathbf{q}) \, q^2 \,.
 \end{gather}
This counter-term comes from the bispectrum contribution and it is a drawback for the information gained from the three-point function\footnote{As pointed out by  \cite{Philcox:2020fqx} for the marked field and analogously for the log-density field, we expect all the terms in the $M^{(n)}$ kernels that are not proportional to $F^{(n)}$ to be exponentially suppressed by filters on the scale $R$. Consequently, $A$ and $\delta$ would have the same UV dependence and the same counter-terms. We however point out in Appendix~\ref{app:scaling} (see Eq.~(\ref{eq:M5newterm})) that one term inside the $M^{(5)}$ kernel is not suppressed by the filtering process and leads to the new counter-term in Eq.~(\ref{eq:pbispec}).}. Therefore, the full set of operators needed at two-loops for $A$ is
\begin{eqnarray}
        \textrm{two-loop operators}:  \{&k^0&, \, k^2, \, P_L, \\ &k^2P_L&,\, P^{1L},\, k^2P^{1L},\, k^4,\,k^4P_L,\, P_{\rm quad},\, k^2 P_{\rm quad},\,P_{\rm bispec} \} \,. \nonumber
\end{eqnarray}

Therefore, there are eleven free parameters needed at two loops. Using the power spectrum for $A$ to fit those parameters demands a lot of attention regarding degeneracy between those terms. Also the number of modes available to fit the theoretical model for $A$ is smaller if compared to $\delta$,\footnote{The fact that the number of modes is in principle smaller does not mean that we have less information for $A$: we expect the four-point function for $A$ to be smaller and as a consequence more information is available in each one of the power spectrum modes of $A$ compared to $\delta$ \cite{Neyrinck:2011xm}.} since the signal is suppressed for $k>R^{-1}$. We thus restrict our analysis in this work to the one-loop EFT corrections.

\section{Comparison to N-body simulations}
\label{sec:results}

In this section we present comparisons between the models described in the last section (Taylor-$A$PT and EoM-$A$PT) and simulation measurements. We also show results using the large-scale bias expansion \cite{Assassi:2014fva,Desjacques:2016bnm} and the power-law fitting function proposed in \cite{Repp:2016gxt}, which we explain later in this section.

We focus more in the applicability of each model to data analysis. Therefore, we only consider the one-loop predictions for the three types of perturbative expansions once this is the order used until now in cosmological analysis \cite{Ivanov:2019hqk,Colas:2019ret, Philcox:2020vvt}.
We also compute the IR-resummed version for each model.

In this work, we do not investigate the constraints of each model in the cosmological parameters. This study must take into account biased tracers and redshift space distortions and it will be the subject of future investigations. 

\subsection{The large-scale bias expansion} \label{sec:bias}

In addition to the methods previously developed in this paper, we also consider the widely used large scale bias expansion to compute the auto power spectrum of $A$ \cite{Assassi:2014fva,Desjacques:2016bnm}. This approach consists of writing down all relevant operators at some order in perturbation theory.
The main advantage of this approach is the broad discussion in the literature on how to apply and compute the theoretical predictions \cite{Nishimichi:2020tvu, CLASSPT}. The idea is to compare the effective field theory model constructed for $A$ on previous sections to an alternative model that considers $A$ as a tracer of $\delta$. Ultimately, testing the bias expansion for $A$ means treating it as a tracer and considering a basis of operator constructed on top of $\delta$. 

In this work we use the public available \texttt{CLASS-PT} \cite{CLASSPT} code in order to compute the auto power spectrum at one loop. This code implements the following expansion for the biased tracer
\begin{equation}
A = b_{1}\delta + \frac{b_{2}}{2}\delta ^{2} + \frac{b_{3}}{6}\delta ^{3} + b_{\mathcal{G}_{2}}\, \mathcal{G}_{2} + b_{\delta \mathcal{G}_{2}}\, \delta  \mathcal{G}_{2} + b_{\mathcal{G}_{3}}\,\mathcal{G}_{3} + b_{\Gamma _{3}} \,\Gamma _{3} + R_{\star}^{2}\, \partial ^{2} \delta + \epsilon \,,
\label{eq:P1l_bias}
\end{equation}
where
\begin{eqnarray}
    \mathcal{G}_{2}(\Phi _{g}) &\equiv& \left( \partial _{i} \partial _{j} \Phi _{g}\right)^{2} - \left( \partial _{i}^{2} \Phi _{g}\right)^{2} \,, \\
    \Gamma _{3} &\equiv& \mathcal{G}_{2}(\Phi _{g}) - \mathcal{G}_{2}(\Phi _{v}) \,,
\end{eqnarray}
with $\Phi _{g}$ and $\Phi _{v}$ being respectively the gravitational and velocity potentials.
The operators $\delta ^{3}$, $\delta \mathcal{G}_{2}$ and $\mathcal{G}_{3}$ are absorbed by renormalization \cite{Assassi:2014fva}. We consider the stochastic term up to second order in derivatives $\epsilon = c_{0} + c_{1}k^{2}$, which together with $b_1$ resemble the free parameters~(\ref{eq:free2}) considered to fix the large-scale problems of perturbation theory in Sec.~\ref{sec:largescales}. 
The final expression for the auto power spectrum of the $A$ field is described by Eq.~(2.10) of \cite{CLASSPT}, which accounts for seven free parameters. 

\subsection{The  power-law fit}
\label{sec:fitting_function}

In addition to all theoretical models described in the previous sections, we also consider a simple power-law fitting function strongly inspired by the one present in \cite{Repp:2016gxt}
\begin{eqnarray}\label{eq:fitting}
    P_A (k) = b_{A}^{2} \,P_{A}^{(11)}(k)\,C(k) \,,
    \label{eq:P_fit}
\end{eqnarray}
where
\begin{eqnarray}
    C(k) = 
    \begin{cases}
    \left(\frac{k}{k_{p}}\right)^\alpha\,,& \text{if } k \geq k_{p}\\
    \,\,\,\,\,\,1\,,              & \text{otherwise} \,.
\end{cases}
\end{eqnarray}
$b_{A}$, $k_{p}$, and  $\alpha$ are free parameters that describe respectively the large-scale bias of $P_{A}$, the scale on which the spectrum is enhanced and the enhancement exponent.

Although simplified, this expression intends to capture the main differences between the $A$ and $\delta$ power spectra. Notice that, aiming for simplicity, we neglected stochastic terms in this power-law fitting function.

\subsection{Effect of the IR resummation}

Before showing the complete result after fitting each model's free parameters, we comment now on the effect of the IR resummation in the final predictions. For both the Taylor-$A$PT and EoM-$A$PT models, we implemented the resummation scheme presented in \cite{Senatore:2014via,Cataneo:2016suz} 
\begin{eqnarray} \label{eq:resummation}
    P^{EFT,1L + IR}_A(k) = \sum_{j=0,1}\sum_{X_j}P_{X_j}(k)||_{1-j},
\end{eqnarray}
in which $P_{X_j}(k)||_{1-j}$ states for each one of the loop terms of order $j$ that appear in the final EFT calculation Eq.~(\ref{eq:eft1L}) and are resummed at order $1-j$. This resummation is done through a convolution of each term with a kernel (see \cite{Senatore:2014via} for a complete expression of the resummation kernel). 
Moreover, we consider the IR-resummation scheme for $A$ to be calculated the same way as for $\delta$, which should be valid at linear order since both tree-level power spectra are the same but for a filter.
Since at non-linear order the BAO wiggles may affect $A$ in a different manner than $\delta$ \cite{McCullagh:2012nu}, a broader study of the BAO resummation effects on $A$ is still necessary.

For the IR-resummation of the large-scale galaxy expansion described in Sec.~\ref{sec:bias}, we used the \texttt{CLASS-PT} implementation\footnote{The resummation in that case is a simplified version of the one considered by \cite{Senatore:2014via} and it is broadly discussed in \cite{CLASSPT}.}.
The IR-resummed version of the power-law fitting function (Sec~\ref{sec:fitting_function}) was computed replacing the linear power spectrum $P_{A}^{(11)}$ by its resummed version at linear order.

\begin{figure}[ht]
\centering  
  \includegraphics[width=0.55\textwidth]{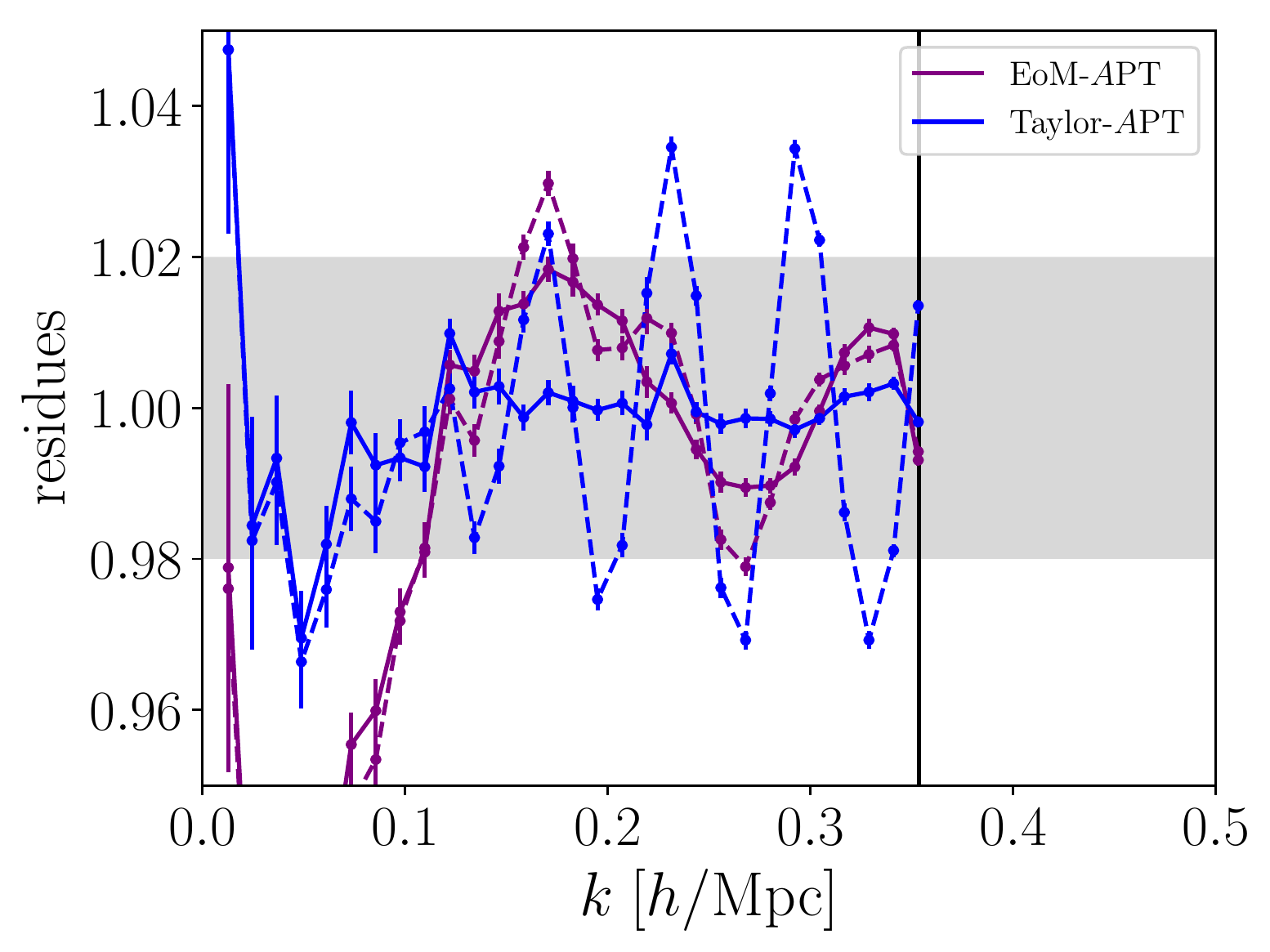}
\caption{\label{fig:resumeffect}
\small Effect of the IR resummation in the residues for both the Taylor-$A$PT and the EoM-$A$PT methods. We show the result before performing the IR resummation by dashed lines and after the IR resummation by solid lines. The black vertical line indicates the smoothing scale $k_{R} = \sqrt{2}\times 0.25$ Mpc$^{-1}$h, which in that case is also the value of $k_{\rm max}$ used for the fit. }
\end{figure}

Fig.~\ref{fig:resumeffect} presents a comparison between the Taylor-$A$PT and the EoM-$A$PT theoretical models for $P_{A}$, with and without the IR-resummation. We display the residues, defined as the ratio between the theoretical prediction and the mean spectrum measured in the N-body simulation set. The results are for the snapshot at $z = 0$ and using $k_{\rm max} \approx 0.35$ Mpc$^{-1}$h for the fits. 
We see that the IR-resummation is more critical for the Taylor-$A$PT model. The reason for that, which we explicitly checked in the fits, is that the one-loop term gives a substantial contribution to this model. The one-loop term is the one that is mostly modified by the resummation scheme at this order in perturbation. We explicitly checked for the EoM-$A$PT model that the resummation is indeed suppressing the BAO wiggles at each term separately. Still, since other terms in the full one-loop expression~(\ref{eq:eft1L}) are dominant (e.g., the stochastic terms), the suppression in that case does not significantly change the final result. 

\subsection{EFT comparison to data}

In order to fit the free parameters of each model, we used the mean $P_{A}(k)$ measured in the 26 N-body simulation realizations and analysed the redshifts $z = 3, \,1, \,0.5$ and $0$. We performed a $\chi ^{2}$ minimization with
\begin{eqnarray} \label{eq:chi2}
    \chi^2 = \frac{\sum_i^{N_{\rm data}}  ( O_i - T_i)^2/\sigma^2_i}{N_{\rm data} - N_{\rm d.o.f.} - 1} \,,
\end{eqnarray}
where $O$ and $T$ are, respectively, the measured power spectrum and the theoretical model. $\sigma^2$ is the standard deviation calculated over the 26 realizations.  

\begin{table}[ht]
    \centering
    \begin{tabular}{l|c | c}
    \hline
    Model & \# of free param. & Equation \\
    \hline
    Taylor-$A$PT & 4(3) & \eqref{eq:eft1L} \\
    EoM-$A$PT & 4(3) & \eqref{eq:eft1L} \\
    Bias exp. & 7(6) & \eqref{eq:P1l_bias} \\
    Power-law. & 3(2) & \eqref{eq:P_fit} \\
    \hline
    \end{tabular}
        \caption{\label{tab:models} {\small Description of each model used in the comparison to N-body simulation data. The second column is the number of free parameters when $b_{1}$ is free (or fixed by $P_\delta$), and the third column indicates the main equation for each model.} }
    
\end{table}

The number of free parameters $N_{\rm free}$ of each model is shown
in Table~\ref{tab:models}. The two different numbers of free parameters correspond to the fits by fixing or not the value of $b_{1}$ through the information of $P_{\delta}$. The fixing of $b_{1}$ was performed using the ratio between $P_{A}$ and $P_{\delta}$ on $k\leq 0.06$ Mpc h$^{-1}$. Note that this procedure does not remove the fitting function's sensitivity on the spectral amplitude $A_{s}$, once this parameter also affects the shape of the power spectrum through the loop corrections.

\begin{figure}[ht]
\centering  
  \includegraphics[width=0.41\textwidth]{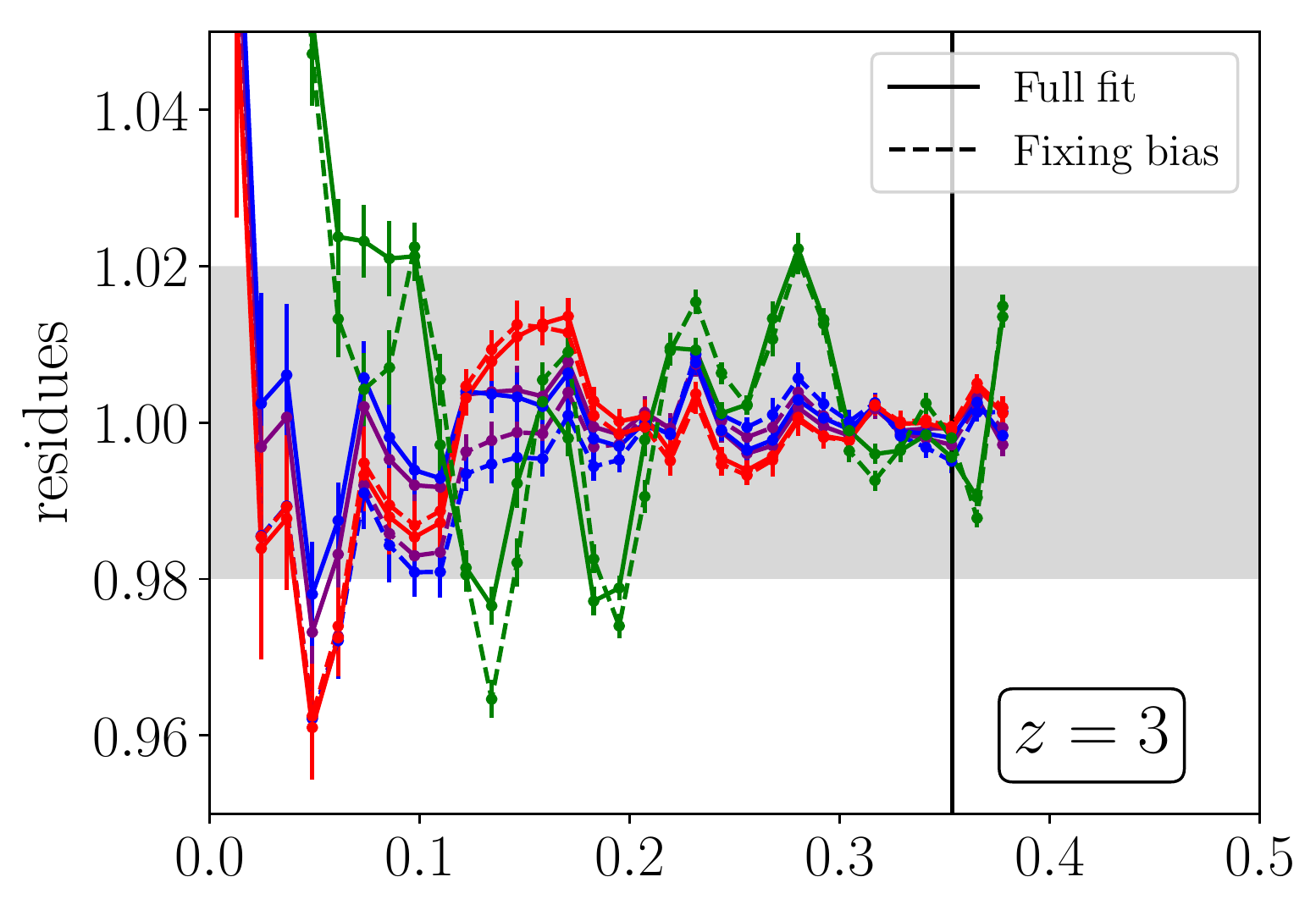}
  \includegraphics[width=0.40\textwidth]{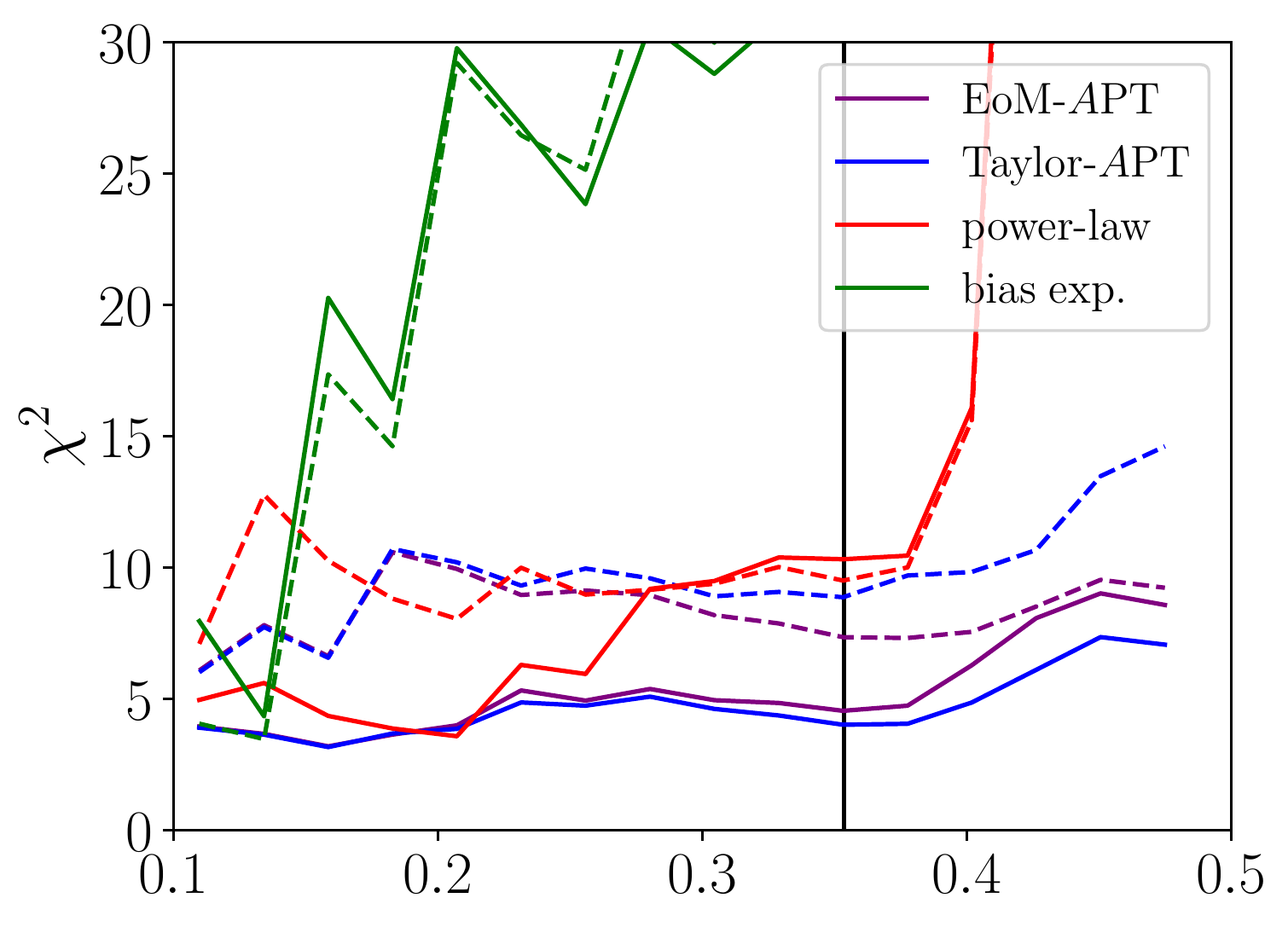}
    \includegraphics[width=0.41\textwidth]{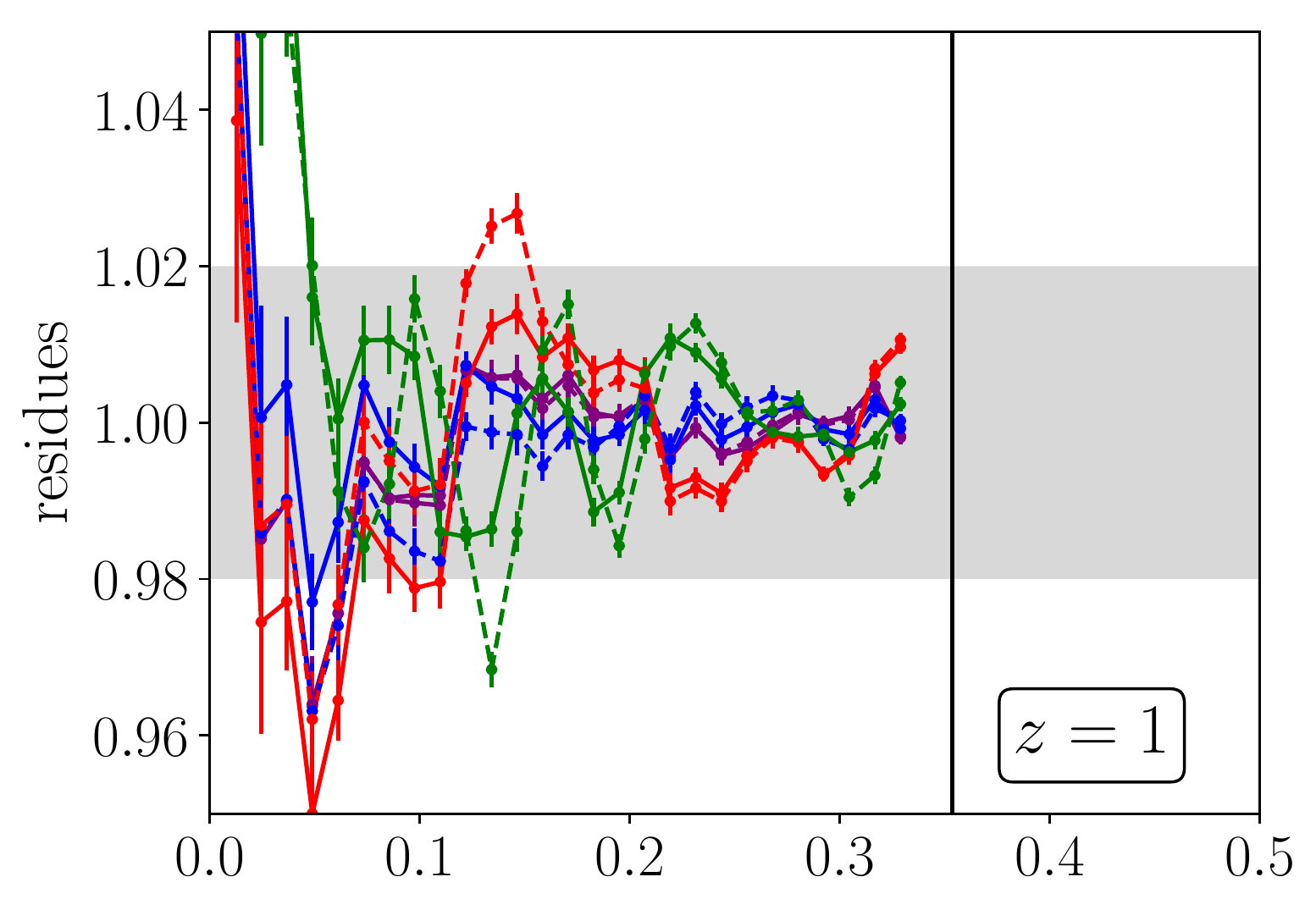} 
  \includegraphics[width=0.4\textwidth]{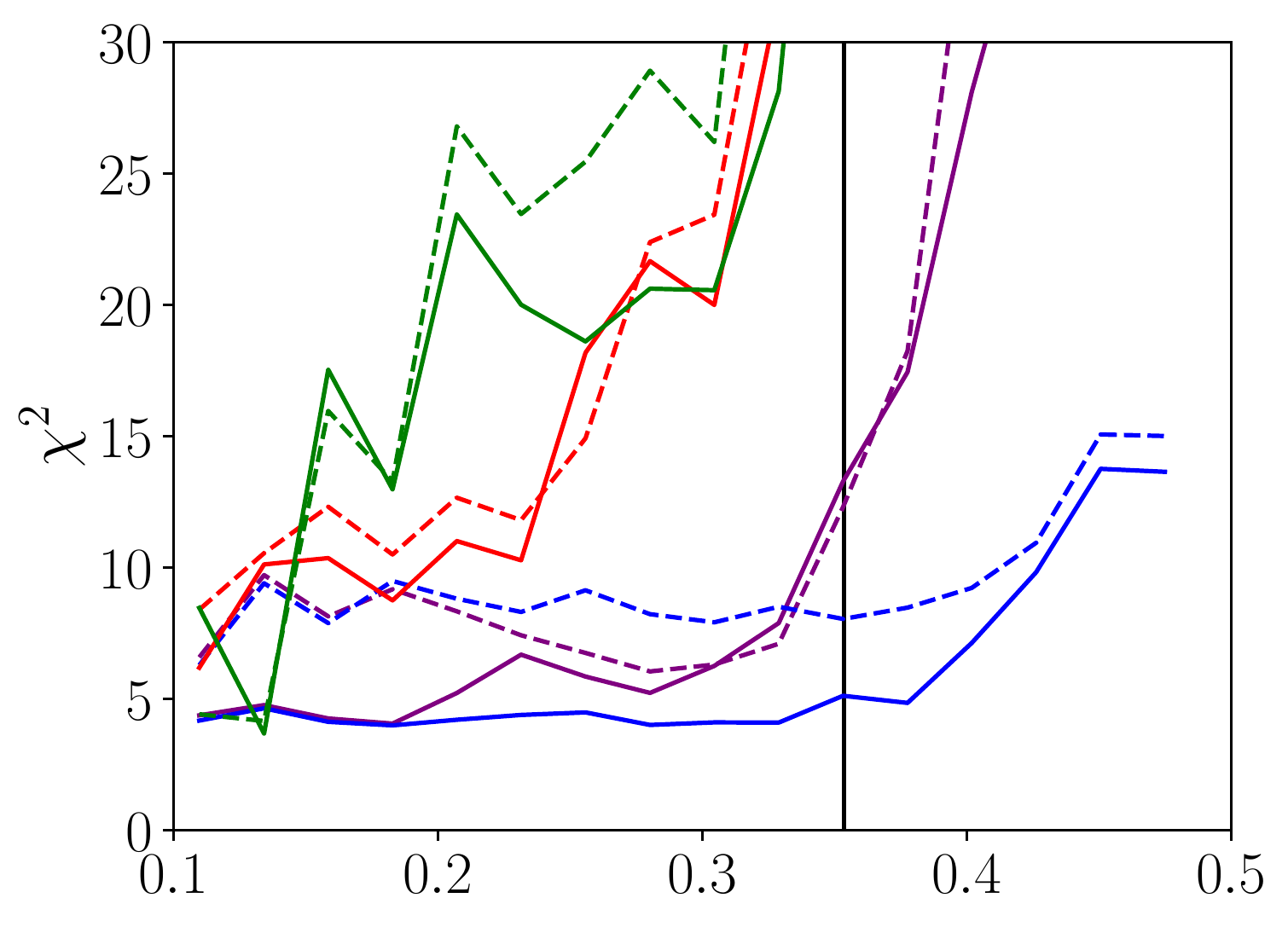}\\
    \includegraphics[width=0.41\textwidth]{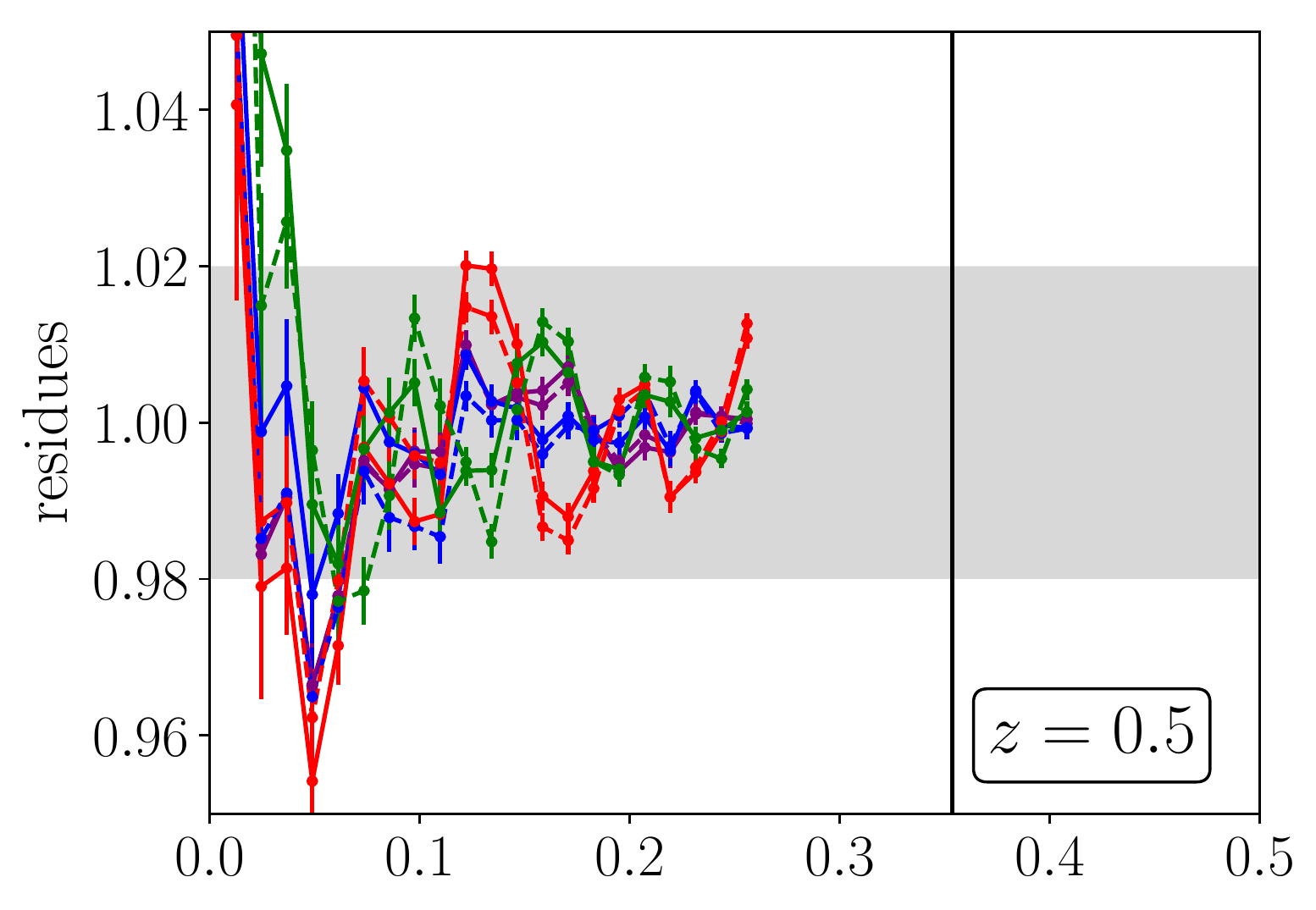} 
  \includegraphics[width=0.4\textwidth]{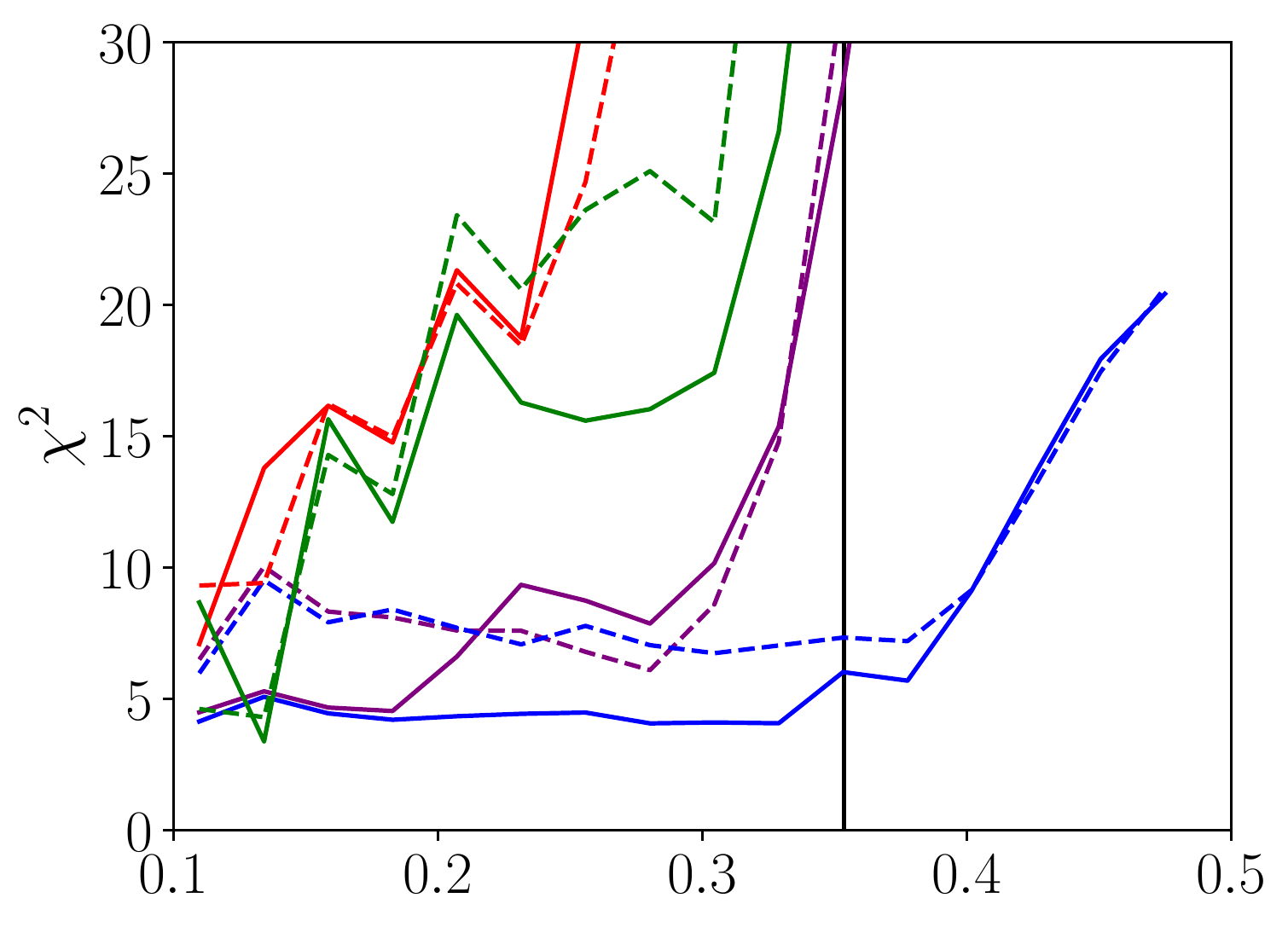}\\
    \includegraphics[width=0.41\textwidth]{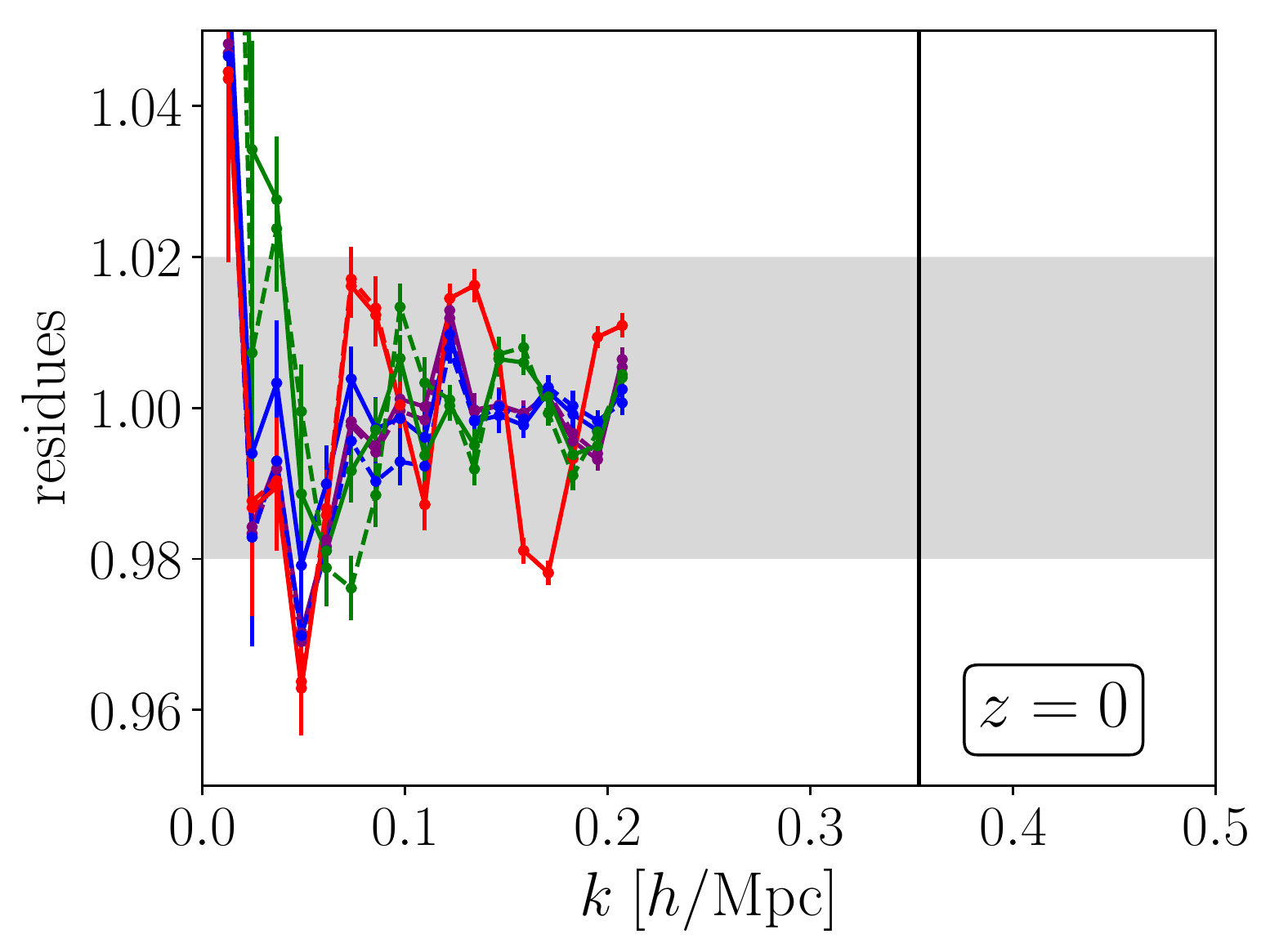}  
  \includegraphics[width=0.40\textwidth]{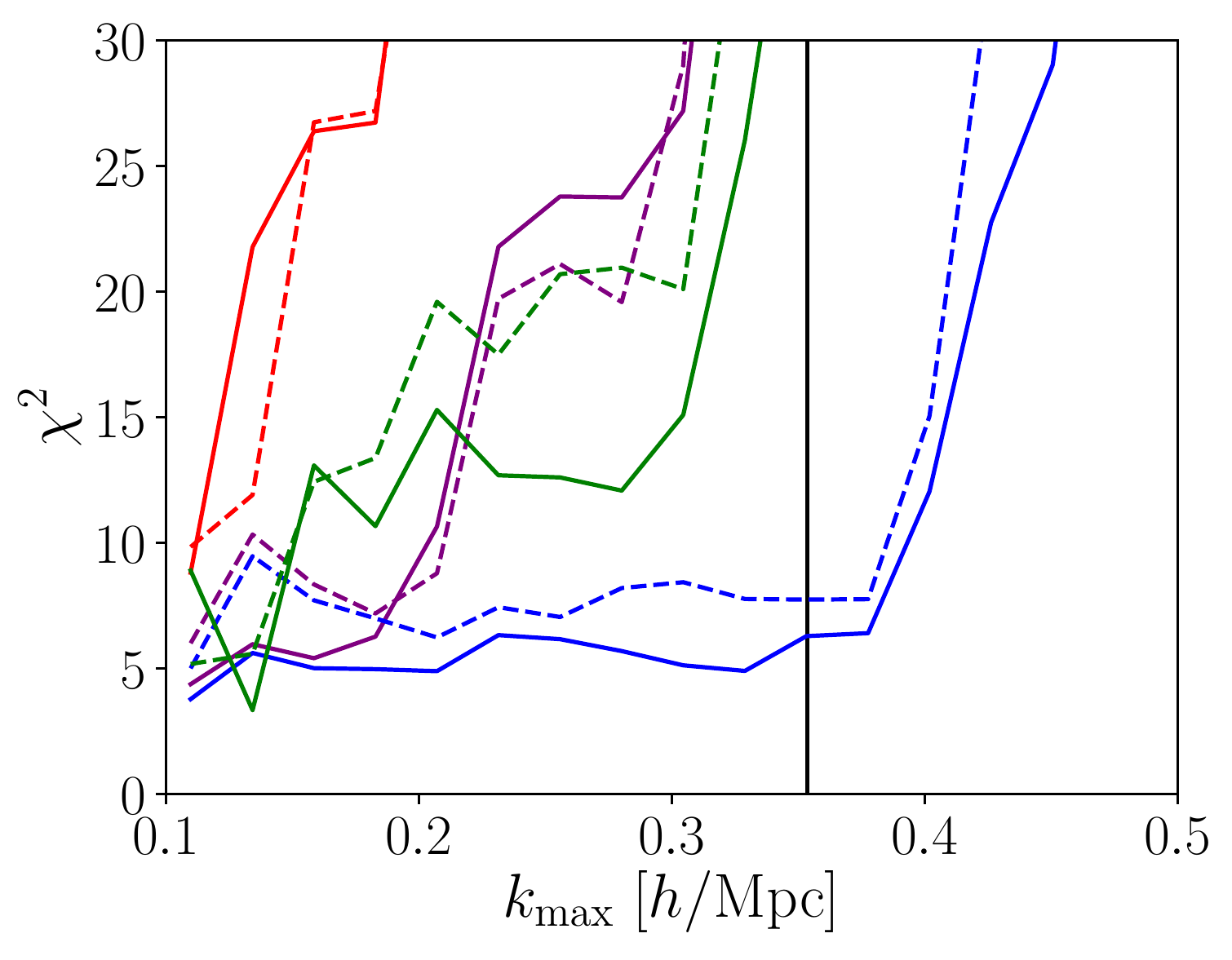} 
\caption{\label{fig:residues}
\small On the left, the residues for the four models described in Table~\ref{tab:models} after fitting the free parameters in simulation data. Each row corresponds to a different redshift, $ z = \{ 3,1,0.5,0\}$. The $k_{\rm max} $ used at each $z$ is indicated by the point where the lines were truncated. On the right, the $\chi^2$ as a function of the $k_{\rm max}$ used for the fit. The solid lines describe the fits of the full set of parameters. The dashed lines correspond to fixing the bias $b_1$ in the larges scales using $P_{\delta}$. The black solid line indicates the smoothing scale $k_R = \sqrt{2}\times 0.25$ Mpc$^{-1}$h in which the spectrum is suppressed.} 
\end{figure}

The left panels of Fig.~\ref{fig:residues} show the residues of each model (different colors), for different redshifts (different rows), and fixing or not $b_{1}$ using $P_{\delta}$ (different line styles). We used for the fits $k_{\rm max}(z) \approx 0.2/D(z)$ Mpc$^{-1}$h, where $D(z)$ is the growth function. $k_{\rm max}$ is also indicated in the figure by the point where the lines are truncated. The right panels show the $\chi ^{2}$ defined according to Eq.~(\ref{eq:chi2}) as a function of $k_{\rm max}$. 
We can see that a $2\%$ accuracy in the residues is achieved by almost all the models at different redshifts, with some offset at small $k$, where the error bars jeopardize the fits.
The worst performance model is the bias expansion. It fails to describe $A$'s power spectrum even at high $z$ despite its large number of free parameters because none of its terms in Eq.~(\ref{eq:P1l_bias}) incorporate information about smoothing scale $R$. It indicates that the most general set of physical operators constructed for $\delta$ is not able to reproduce the new functional dependence of $A$. Notice that Eq.~(\ref{eq:P1l_bias}) does not know about the filtering procedure and, as explained along the text, the filtering affects all the $k$ modes of $A$.
All other models work almost equally well at high $z$ until very high $k_{\rm max}$, since their linear spectrum  with the filter describe the theory pretty well. 
 
 \begin{figure}[ht]
\centering  
  \includegraphics[width=0.48\textwidth]{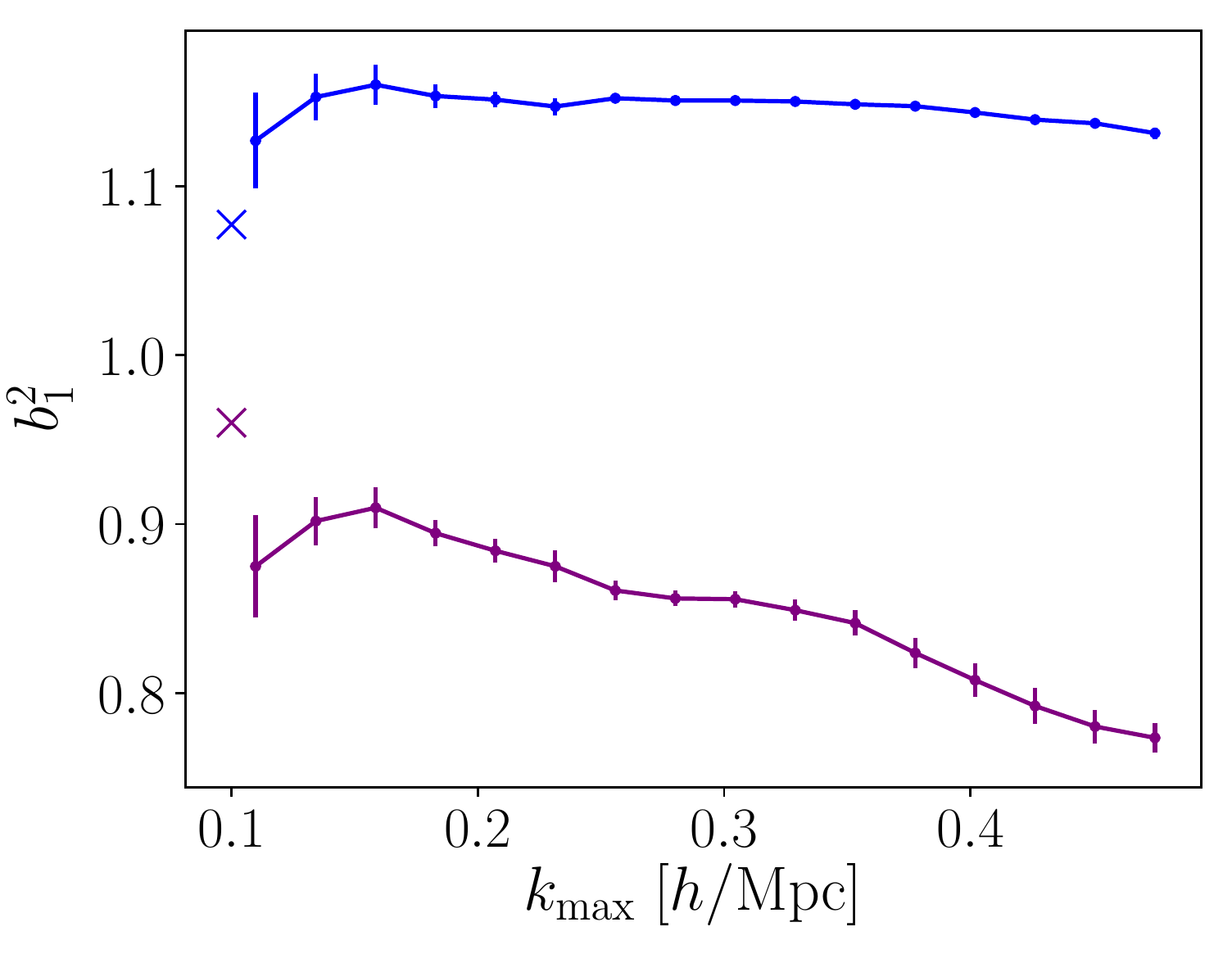}
  \includegraphics[width=0.49\textwidth]{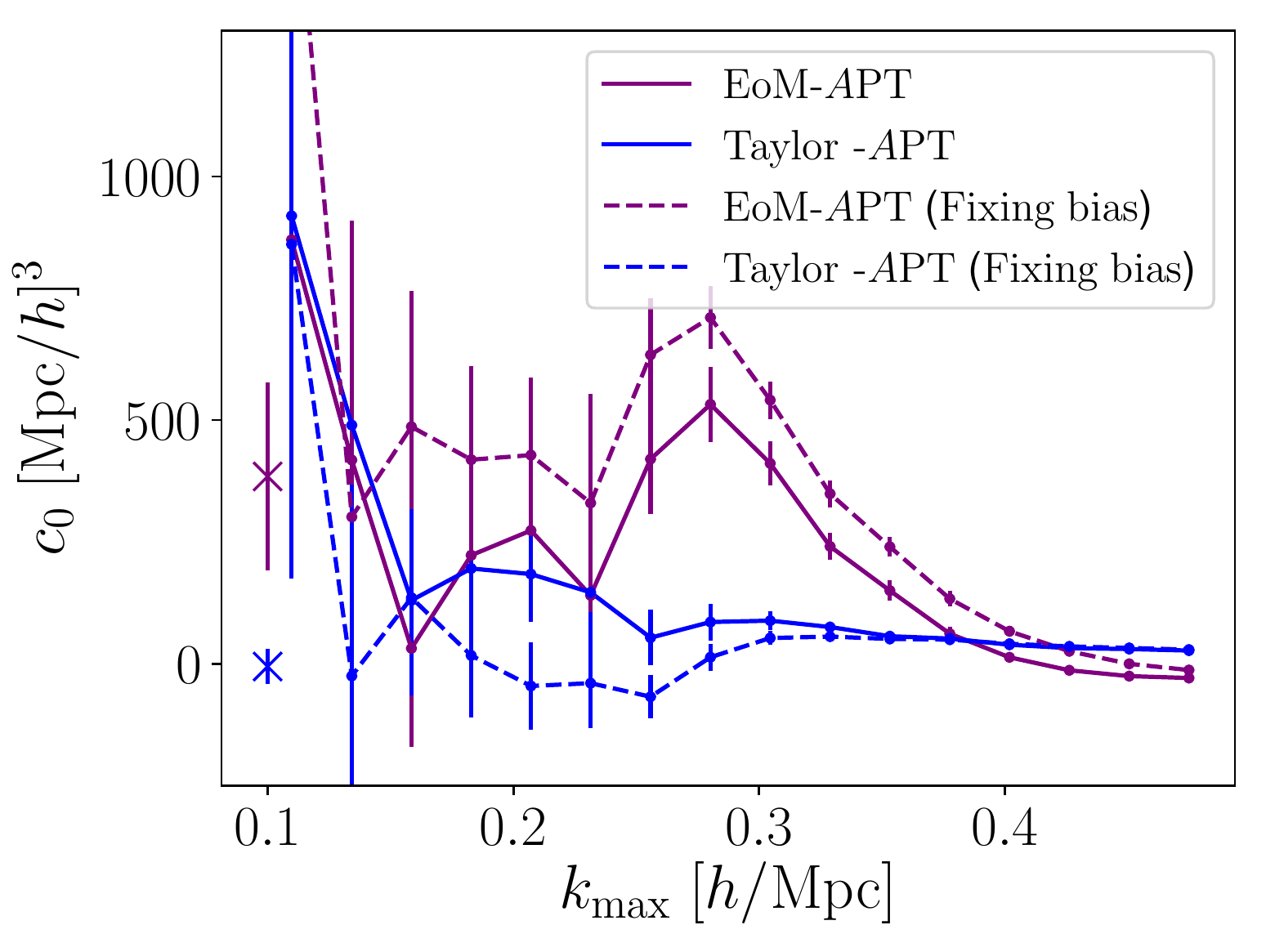}
    \includegraphics[width=0.49\textwidth]{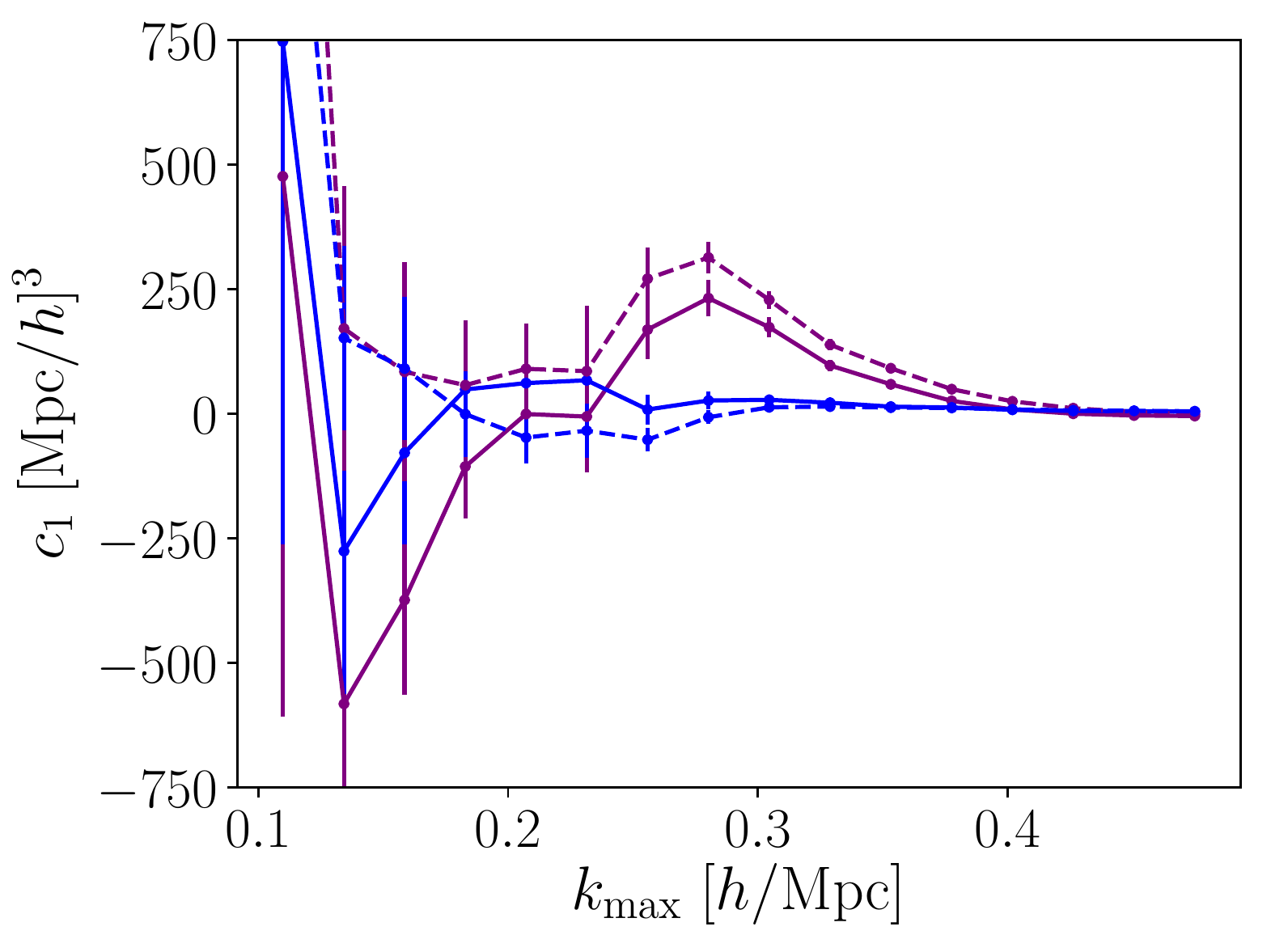} 
  \includegraphics[width=0.49\textwidth]{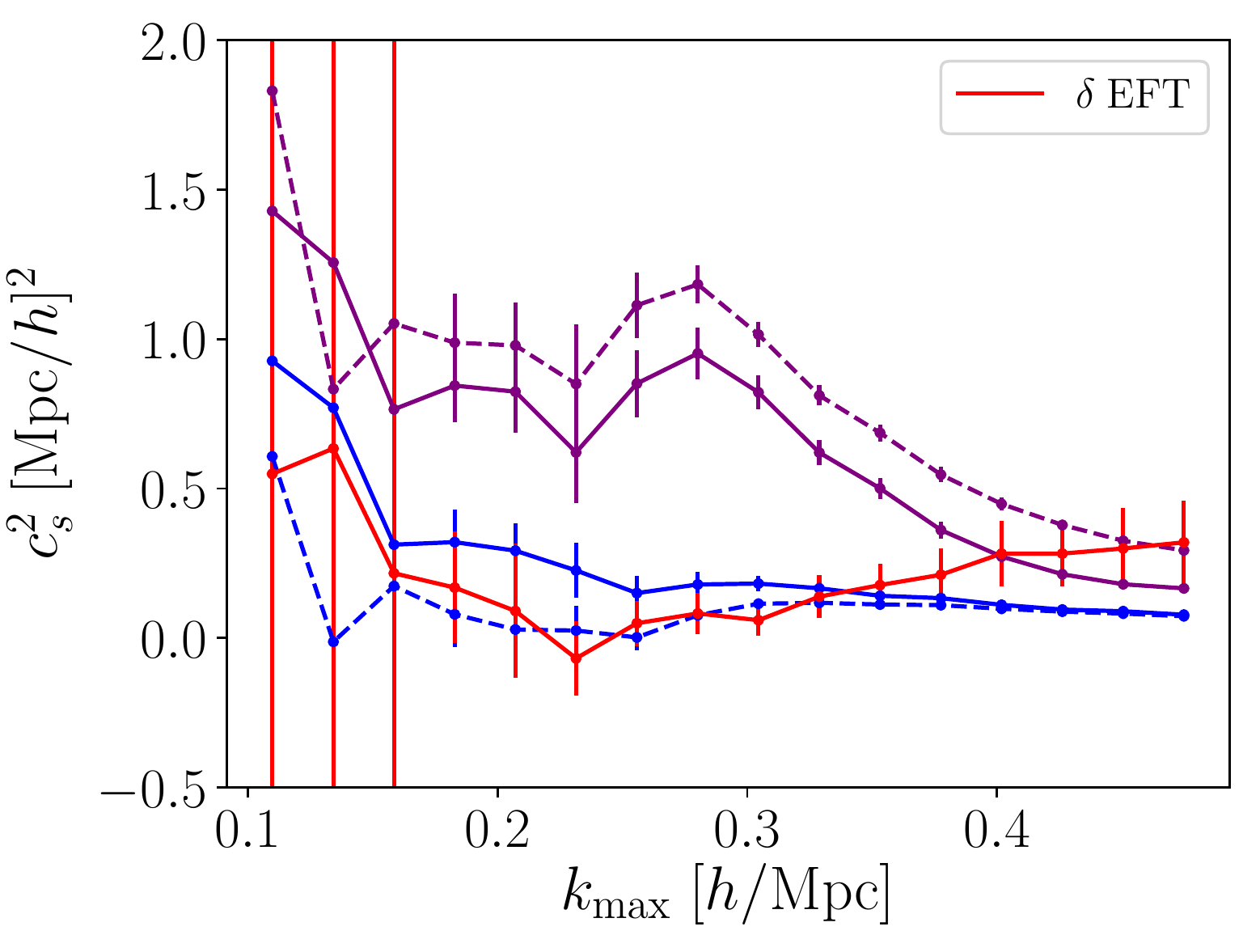} 
\caption{\label{fig:rgflow}
\small Running of the free parameters with $k_{\rm max}$ at $z=0$ for the two models proposed in this work for $A$ and the usual EFT for $\delta$ (where $c_{s}$ is the only free parameter). The blue lines are for the Taylor-$A$PT model, the purple lines for the EoM-$A$PT model, and the red line for the EFT for $\delta$. Solid lines indicate fits with the four free parameters of Eq.~\eqref{eq:eft1L} and dashed lines are the fits keeping $b_{1}$ fixed through the information in the $\delta$ power spectrum. The crosses in the top panels are the best-fit parameters coming from the propagator (as shown in Fig.~\ref{fig:propagator}). } 
\end{figure}

Notice by the difference between solid and dashed lines that fixing the bias on the largest scales through $P_\delta$ (dashed lines) does not lead to large deviations from the N-body data, suggesting that this procedure can be used to reduce the number of free parameters. 
It indicates that using $\delta$ can improve the predictions for the $A$ spectrum. It brings an exciting avenue to study: how combining both probes can improve the cosmological constraints.

The model with the best residues at all $z$ is the Taylor-$A$PT, followed by the EoM-$A$PT. The Taylor-$A$PT scheme leads to considerably better predictions than the bias expansion, even though the last has more free parameters.  The EoM-$A$PT performs equally well at $z=0$ for $k_{\rm max} < 0.2$ Mpc$^{-1}$h, but since the theory described in Sec.~\ref{sec:Apt} manifestly has a cutoff at $R^{-1}$ (black vertical line) it starts to diverge when approaching that scale, as it is expected from effective field theories\footnote{The $R^{-1}$ cutoff does not appear at high $z$ because the linear theory succeeds in describing the measured spectrum. For the Taylor-$A$PT scheme the scale $R$ is not a cutoff but only a scale where some of its terms are suppressed.}. As shown in Sec.~\ref{sec:theory}, the EoM-$A$PT theory performs better on the largest scales, but the addition of the free-parameters~(\ref{eq:plarge}) make both  EoM-$A$PT and Taylor-$A$PT to perform equally well on those scales. 

In Fig.~\ref{fig:rgflow} we show the running of the free-parameters as a function of $k_{\rm max}$ for $z=0$. Notice that while the EoM-$A$PT model presents a strong running of the parameters for $k_{\rm max} > 0.3$ Mpc$^{-1}$h, which corroborates with the argument given above, the Taylor-$A$PT model is stable over all $k_{\rm max}$. This stability indicates that the Taylor-$A$PT model at one-loop can absorb the correct scaling of the power spectrum of $A$ without any evidence of over-fitting. Besides that, when comparing $c_s^2$ for the Taylor-$A$PT when fixing the linear bias (blue-dashed), we can see that it follows the $c_s^2$ obtained by using the usual EFT for $\delta$ (red solid) up to $k_{\rm max} \sim 0.3$ Mpc$^{-1}$h. From that scale on, the counter-term value for the $\delta$ theory starts to deviate, while the $c_s^2$ obtained for $A$ is almost constant. It indicates that the two-loop contribution for $\delta$ is much more relevant on those scales than for $A$, which resembles the expectation that $A$ is more linear. In the case of $\delta$, there are evidences that the two-loop contribution starts to kick in for $k > 0.2 $ Mpc$^{-1}$h \cite{Nishimichi:2020tvu}. Moreover, the crosses at $k_{\rm max} = 0.1$ Mpc$^{-1}$h on the top panels indicate the values obtained for the fit of the propagator in Fig.~\ref{fig:propagator}, which are within $2\sigma$ from the values obtained for the fit to the power spectrum.

\begin{figure}[ht]
\centering
  \includegraphics[width=0.7\textwidth]{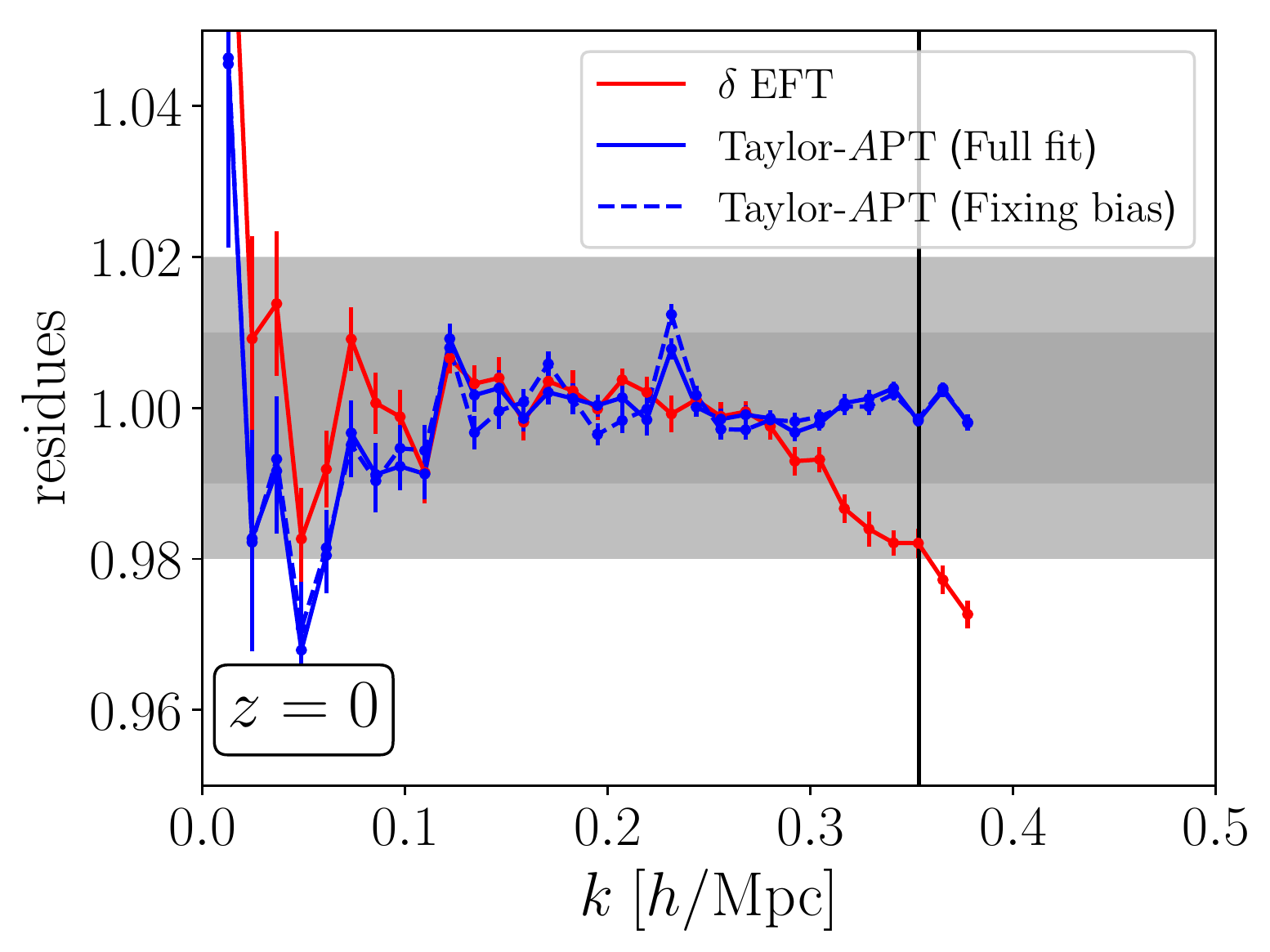}
\caption{\label{fig:moneyplot}%
\small Same as the left panels of Fig.~\ref{fig:residues}, but now using $k_{\rm max} \simeq 0.38 $ Mpc$^{-1}$h at $z=0$. The Taylor-$A$PT model can reach very non-linear scales with precision of $1\%$ and fixing the bias parameter (dashed) does not jeopardize the theory.}
\end{figure}

When including the free-parameters~(\ref{eq:plarge}) together with the EFT counter-term~(\ref{eq:ct1}) the Taylor-$A$PT expansion is enough to push the theoretical limits until high values of $k_{\rm max}$, namely $k_{\rm max} \simeq 0.38 $ Mpc$^{-1}$h at $z=0$. 
We show in Fig.~\ref{fig:moneyplot} the comparison between the residues for the one-loop effective field theory for $\delta$ and for $A$ using the Taylor expansion at $z=0$ and taking $k_{\rm max} \simeq 0.38 $ Mpc$^{-1}$h. The higher number of free parameters for $A$ and specially the $k^2$ term make the EFT for $A$ slightly better for $k \simeq 0.35 $ Mpc$^{-1}$h. Notice that those new terms are already present in the large-scale bias expansion and they will not introduce any new free parameter in a realistic cosmological analysis where biased tracer and redshift space distortions are taken into account. As long as no new degeneracies are induced between the nuisance parameters and the cosmological ones it will not jeopardize the physical constraints.
Further investigation is still needed to quantify how much information about cosmological parameters can be extracted from the EFT for $A$ if compared to $\delta$ and how to consistently use both fields in a jointly analysis.

\section{Conclusions}
\label{sec:conclusion}

In this work we calculated the perturbative series and its effective field theory corrections for the logarithm-transformed density field. We have shown that the series is asymptotic already on the large scales when the perturbative approach is constructed on top of a Taylor expansion of the log function. Adding more loop corrections jeopardizes the theoretical prediction for small $k$ instead of improving it. Different than for $\delta$, the large scale prediction for the log density spectrum receives relevant contributions from intermediate scales $k<R^{-1}$ \cite{Neyrinck:2013iza} where the perturbation theory for $\delta$ has evidence to be asymptotic \cite{Blas:2013aba,Konstandin:2019bay}. We have constructed an alternative perturbative approach for $A$ directly through the equations of motion and this scheme has reproduced the large-scales with better accuracy at $z=0$. This new approach also has a better hierarchical structure between each loop correction.

Afterwards, we have shown that the large scales are fixed by adding two free parameters~(\ref{eq:free}) in both the Taylor-$A$PT and the EoM-$A$PT schemes at tree level. Those free parameters were already considered in the case of the marked fields in \cite{Philcox:2020srd}. We have shown that the inclusion of those parameters -- at least in the log transformation case -- not only takes into account the resummation of higher-order terms but also fixes the series' convergence issues described above in the text. We have also calculated the counter-terms that parametrize the UV in the effective field theory formulation by analysing the UV scaling of each loop diagram. An additional counter-term appears at two-loops if compared to the EFT for $\delta$. This new term comes from the $\delta$'s bispectrum and is somewhat a price to pay for the information brought from the three-point function of $\delta$ down to the two-point function of $A$. 

Finally, we compared the effective field theory prediction for $A$ to an N-body simulation suite. We also considered two other theories for $A$ in addition to the models described in Sec.~\ref{sec:theory}: a power-law fitting function based on \cite{Repp:2016gxt} and the large-scale bias expansion \cite{Desjacques:2016bnm}. We also implemented the IR-resummation based on \cite{Senatore:2014via} showing that it improves the Taylor-$A$PT by removing the wiggles.  

Furthermore, when comparing the models to data, we found that the Taylor-$A$PT scheme outperforms the others and can extend the theoretical predictions until $k \simeq 0.38 $ Mpc$^{-1}$h. The EoM-$A$PT scheme performed well on scales $k<0.2 $ Mpc$^{-1}$h and fails when getting closer to $R^{-1}$. It agrees with the behaviour expected from an effective field theory since this scale is a manifest cutoff of that scheme. Besides that, the running of $c_s^2$ within the Taylor-$A$PT model  agrees with the value obtained by the EFT formulation for $\delta$ up to $k\sim 0.3 $ Mpc$^{-1}$h. The value of $c_s^2$ obtained by $\delta$ changes for higher $k$, while it remains constant for the Taylor-$A$PT. It indicates that the two-loop contribution for $A$ might only kick in on smaller scales, which resembles the expectation that the log-transformed field is more linear.

We have shown that the effective field theory formulation for $A$, together with a couple of extra free parameters that fix the large scales, enable us to extend the theoretical prediction for that field until non-linear scales.  It is vital to notice that the goal of extracting information out of $A$ is not orthogonal to use the standard EFT prediction for $\delta$. For instance, we have shown that using information from $\delta$ can reduce the number of free-parameters for $A$. An exciting goal in the short horizon is to study how combining both probes can improve the cosmological parameters constraints.

\acknowledgments
We are grateful to Thomas Konstandin, Oliver Philcox and Fabian Schmidt for providing valuable comments on the manuscript.
We are thankful to the MPA infrastructure. The N-body simulations were run in MPA's FREYA cluster.
RV is supported by FAPESP. 
HR is supported by the Deutsche Forschungsgemeinschaft under Germany's Excellence Strategy - EXC 2121 "Quantum Universe" - 390833306.

\appendix

\section{Scaling of the the PT terms} \label{app:scaling}

In this appendix we calculate the UV scaling of the PT terms for $\delta$ and $A$. We consider the single-hard and double-hard limits of the one and two loops PT integrals in order to check the cutoff ($\Lambda$) dependence. We assume a universe scaling in the UV as $P_L(k) \propto k^n$.

\subsection{For $\delta$}
We start by calculating the leading order contribution of the loop integrals for $\delta$ in the same way as done in \cite{Baldauf:2015aha}. We use the  limit of the kernels when you have a hard mode but the total sum $\mathbf{k} = \sum \mathbf{q}_n$ is kept constant (see \cite{Goroff:1986ep,Bernardeau:2001qr})
\begin{eqnarray} \label{eq:Flimit}
      F^{(n)}(\mathbf{q}_1,\dots,\mathbf{q}_{n-2},\mathbf{p},-\mathbf{p})     &\overset{p\rightarrow\infty}{\propto}& \left( \frac{k}{p}\right)^2 , \left( \frac{k}{p}\right)^4  ,\, \dots  
\end{eqnarray}

At one loop, the leading order contributions of each term in the limit $q\gg k$  is given by 
\begin{eqnarray} 
          P_{\delta}^{(13)}(k) &\overset{q\rightarrow\infty}{\propto}& P_L(k) \int_\mathbf{q}  \ P_L(q) \left(\frac{k}{q}\right)^2 
          \propto 
          k^2 P_L(k) \Lambda^{n+1} 
          \,,  \label{eq:p13deltaUV}
\\
          P_{\delta}^{(22)}(k) &\overset{q\rightarrow\infty}{\propto}&  \int_\mathbf{q}  \left(\frac{k}{q}\right)^4 \left[ P_L(q)\right]^2  \propto k^4 \Lambda^{2n-1} \,.\label{eq:p22deltaUV}
\end{eqnarray}
Analysing the divergence structure, we can conclude that, at one loop, the counter-term needed to cancel out the $\Lambda$ dependence is 
\begin{eqnarray} \label{eq:delta1ct}
    P_{\rm c.t.}(k) = c_s^2 k^2P_L(k)\,.
\end{eqnarray}

We start the two-loop analysis by calculating the single-hard limit, in which one of the two momenta is larger than the other
\begin{eqnarray}
              P_{\delta}^{(15)}(k) &\overset{q_1\rightarrow\infty}{\propto}&  P_L(k) \int_{\mathbf{q}_{1}} \left(\frac{k}{q_1}\right)^2 P_L(q_1) \propto  k^2 P_L(k)\Lambda^{n+1}  \,,
              \\
            P_{\delta}^{(24)}(k) &\overset{q_1\rightarrow\infty}{\propto}& \int_{\mathbf{q}_{1}} \left(\frac{k}{q_1}\right)^4 \left[P_L(q_1)\right]^2 \propto k^4 \Lambda^{2n-1}  \,,  
            \\
            P_{\delta}^{(24)}(k) &\overset{q_2\rightarrow\infty}{\propto}& \int_{\mathbf{q}_{12}} F^{(2)}(\mathbf{k}-\mathbf{q}_1,\mathbf{q}_1)P_L(q_1) \left(\frac{k}{q_2}\right)^2 P_L(q_2) 
            \\
            &\propto& k^2 \Lambda^{n+1}\int_{\mathbf{q}_{1}} F^{(2)}(\mathbf{k}-\mathbf{q}_1,\mathbf{q}_1)P_L(q_1) \,,  
            \nonumber \\
        P_{\delta}^{(33a)}(k)  &\overset{q_1\rightarrow\infty}{\propto}&   P_L(k)\int_{\mathbf{q}_{12}}F^{(3)}(\mathbf{k},\mathbf{q}_2,-\mathbf{q}_2) P_L(q_2) \left(\frac{k}{q_1}\right)^2 P_L(q_1) 
        \\
        &\propto& k^2 P_L(k) P_{\delta}^{(13)}(k)\Lambda^{n+1}  \,, 
           \nonumber \\
    P_{\delta}^{(33b)}(k) &\overset{q_1\rightarrow\infty}{\propto}&  \int_{\mathbf{q}_{1}}\left(\frac{k}{q_1}\right)^4 \left[P_L(q_1)\right]^2\propto k^4 \Lambda^{2n-1}   \, .
\end{eqnarray}
Now the double-hard limit ($q_1,q_2 \gg k$)
\begin{eqnarray}
    P_{\delta}^{(15)}(k) &\overset{q_1,q_2\rightarrow\infty}{\propto}&  P_L(k) \int_{\mathbf{q}_{12}} \left(\frac{k}{q_{12}}\right)^2 P_L(q_1)P_L(q_2) \propto  k^2 P_L(k)\Lambda^{2n+4} \,, 
    \\
    P_{\delta}^{(24)}(k) &\overset{q_1,q_2\rightarrow\infty}{\propto}&  \int_{\mathbf{q}_{12}} \left(\frac{k}{q_{1}}\right)^2\left(\frac{k}{q_{2}}\right)^2 \left[P_L(q_1)\right]^2P_L(q_2) \propto  k^4 \Lambda^{3n+2} \,,
    \\    
    P_{\delta}^{(33a)}(k) &\overset{q_1,q_2\rightarrow\infty}{\propto}& P_L(k) \int_{\mathbf{q}_{12}} \left(\frac{k}{q_{1}}\right)^2\left(\frac{k}{q_{2}}\right)^2 P_L(q_1)P_L(q_2)  \propto  k^4 P_L(k)  \Lambda^{2n+2} \,, 
    \\
    P_{\delta}^{(33b)}(k) &\overset{q_1,q_2\rightarrow\infty}{\propto}&  \int_{\mathbf{q}_{12}} \left(\frac{k}{q_{12}}\right)^4 P_L(q_{12})P_L(q_1)P_L(q_2) \propto  k^4 \Lambda^{3n+2} \, .
\end{eqnarray} 
As a consequence, the counter-terms that cancel out the two-loop divergences are $k^2P_L$, $k^4$ and $k^4P_L$. Notice that the single-hard limit for $P_{\delta}^{(33a)}$ demands us to include a term proportional to $k^2P_{\delta}^{(13)}$, in that case $k^2P^{1L}$. The single-hard limit of $P_{\delta}^{(24)}(k)$ also introduces a divergence that comes from the bispectrum and is canceled by a counter-term like
\begin{gather} \label{eq:pquad}
  k^2 P_{\rm quad} = k^2 \int_\mathbf{q}  P_L(|\mathbf{k}-\mathbf{q}|)P_L(q)F^{(2)}(\mathbf{k}-\mathbf{q},\mathbf{q})\,.
 \end{gather}
Therefore, the full set of counter-terms at two loops is 
\begin{eqnarray} \label{eq:delta2L}
    \{ k^2P_L, k^2P^{1L}, k^4,k^4P_L, k^2 P_{\rm quad} \} \,.
\end{eqnarray}

\subsection{For $A$}

We now consider the hard limit for $A$. We focus on the Taylor-$A$PT framework described in Sec.~\ref{sec:deltapt} because its kernel structure is manifestly evident. As explained in the main text, the EoM-$A$PT case resembles the Taylor-$A$PT in the limit $W\rightarrow1$, such that all counter-term structures are similar but for a filter. 

 Before calculating the limits, one comment is in order. Notice that the limit of $F$ described by Eq.~(\ref{eq:Flimit}) is only valid when the total sum of the arguments is kept constant. For the M kernels, this is not always the case. For instance 
\begin{eqnarray} \label{eq:M3limit}
    M^{(3)}(\mathbf{k},\mathbf{q},-\mathbf{q}) \supset F^{(2)}(\mathbf{k},\mathbf{q}) &\overset{q\rightarrow\infty}{\propto}& \frac{q}{k} \,.
\end{eqnarray}
This limit can be seen by the $F^{(2)}$ explicit form \cite{Bernardeau:2001qr} and it was numerically tested for higher-order kernels. In practice, those non-local  divergences are odd in $q$ and always vanish when integrated. Therefore, the only terms that contribute for each $M^{(n)}(\mathbf{q}_1,\dots,\mathbf{q}_{n}) $ kernel in the hard limit are those ones whose the total sum $\mathbf{k} = \sum \mathbf{q}_n$ is kept fixed.

Moving on to the calculation of the single-hard modes for each one of the kernels $M$, defined in Eqs.~(\ref{eq:M1})-(\ref{eq:M5}), we find
\begin{eqnarray}
    M^{(2)}(\mathbf{k}-\mathbf{q},\mathbf{q}) &\overset{q\rightarrow\infty}{\propto}&  \Bigg\{ W(kR)\frac{k^2}{q^2}  , \, W^2(qR) \Bigg\} \,,
 \\
    M^{(3)}(\mathbf{k},\mathbf{q},-\mathbf{q}) &\overset{q\rightarrow\infty}{\propto}&  \Bigg\{ W(kR)\frac{k^2}{q^2}  , \, W^2(qR), W^2(qR)W(kR) \Bigg\} \,.
\end{eqnarray}
Replacing in the spectra we have
\begin{eqnarray} 
     P_{A}^{(13)}(k) &\overset{q\rightarrow\infty}{\propto}&  W(kR) P_L(k) \int_\mathbf{q}  P_L(q) \left[ W(kR) \left( \frac{k}{q}\right)^2  ,  \, W^2(qR),\, W^2(qR)W(kR)\right] 
      \nonumber \\
     &\propto&  k^2 W^2(kR) P_L(k)    \Lambda^{n+1} ,\,  W(kR) P_L(k)e^{-R\Lambda} ,\,  W^2(kR) P_L(k)  \, e^{-R\Lambda} \,, \label{eq:p13_uv_a}
     \\
    P_{A}^{(22)}(k) &\overset{q\rightarrow\infty}{\propto}&  \int_\mathbf{q} \left[ W(kR) \frac{k^2}{q^2} , \, W^2(qR) \right]^2 \left[P_L(q)\right]^2\,,
        \nonumber \\
     &\propto& k^4W^2(kR) \Lambda^{2n-1} , \, k^2W(kR) \, e^{-R\Lambda} ,\,  e^{-R\Lambda}  \,,\label{eq:p22_uv_a}
\end{eqnarray}
where we used $e^{-R\Lambda}$ to refer to terms that are exponentially suppressed (not necessarily with this strict exponential dependence). We highlight some important points regarding the hard limit of those diagrams:
\begin{itemize}
    \item We spot two kinds of UV contributions. First, those that depend on $\Lambda$ to some power which are already present in the calculation for $\delta$ (see Eqs.~(\ref{eq:p13deltaUV}) and (\ref{eq:p22deltaUV})) and will be canceled out by a counter-term as Eq.~(\ref{eq:delta1ct}). Second, those terms that are suppressed by $R$. The terms suppressed by $R$ come from the $F$ independent part of the $M$ kernels. Those terms also add a finite contribution before suppression. There is therefore a dependence on $R$ that is physical, affects all the scales and is parametrized by $\epsilon_{\rm shot}$ and $\sigma^2$ defined in Eqs.~(\ref{eq:eps}) and  (\ref{eq:variance}). As can be seen by Eqs.~(\ref{eq:p22_uv_a}) and (\ref{eq:p13_uv_a}), the physical contribution of those terms are proportional to 
    \begin{equation} \label{eq:lowkcontrib}
        \{ k^0, k^2, P_L\} \,.
    \end{equation}
    Since the constant $k^0$ works as a shot noise term, we label it as $P_{\rm shot}$.
    
    \item As argued in Sec.~\ref{sec:largescales}, even though the contributions in Eq.~(\ref{eq:lowkcontrib}) are physical they fail to reproduce the low-$k$ behaviour of the PT series (mainly for the Taylor-$A$PT scheme). As shown in the same section, it is not only a matter of including higher-loop contributions or fixing the propagator, but the series is not convergent even on small-$k$. Consequently, we need to add three free parameters proportional to the terms in Eq.~(\ref{eq:lowkcontrib}) in order to fix the large scales. We make a language distinction and call them  free-parameters and not counter-terms since they do not fix the UV scaling.
    
    \item Notice that the exponential suppression only happens when using a Gaussian filter. For other kinds of filter, as for instance a top hat in position space considered in other works \cite{Neyrinck:2009fs}, the suppression is only polynomial in $R\Lambda$. It leads to non-negligible contributions for modes $q > R^{-1}$ or even modes already in the non-linear regime. In that case, the free-parameters in Eq.~(\ref{eq:lowkcontrib}) would also start to play a role of counter-terms and will absorb important UV contributions. 
    
    \item Different powers of $W(kR)$ might appear in the UV analysis, e.g. $W(kR) P_L$ together with $W^2(kR) P_L$. Since we are typically interested on scales $k < R^{-1}$, it is a good approximation to absorb them into the $W^2(kR)$ term (e.g. $W^2(kR) P_L$).
\end{itemize}

Therefore, the set of free parameters and counter-terms needed to cancel out the divergences at one-loop are
\begin{eqnarray}
    \{ k^0, k^2, P_L, k^2P_L \} \,.
\end{eqnarray}

Continuing to the two-loop calculation, we find for the single-hard limits\footnote{As seen for $\delta$, the double-hard limit does not bring any new counter-term, but only leads to different scaling with the cutoff. We therefore restrict our analysis here to the single-hard limit. }  
\begin{eqnarray}
    &M^{(3)}(\mathbf{k}-\mathbf{q}_1-\mathbf{q}_2,\mathbf{q}_1,\mathbf{q}_2)  \overset{q_1\rightarrow\infty}{\propto} \\
   & \Bigg\{
    W(kR) \frac{k^2}{q_1^2} , W(kR)\frac{k^4}{q_1^4} ,
     \,  W( |\mathbf{k} - \mathbf{q}_2| R)W(q_2)\frac{|\mathbf{k} - \mathbf{q}_2|^2}{q_1^2}
    ,  \, W^2(q_1R)W(q_2R) \Bigg\} \nonumber \,, \\
     &   M^{(4)}(\mathbf{k}-\mathbf{q}_1,\mathbf{q}_1,\mathbf{q}_2,-\mathbf{q}_2)  \overset{q_1\rightarrow\infty}{\propto} \Bigg\{
    W(kR) \frac{k^2}{q_1^2} , W(kR)\frac{k^4}{q_1^4} , 
    \\
    & \,W(|\mathbf{k}+\mathbf{q}_2|R)W(q_2 R)\frac{|\mathbf{k} + \mathbf{q}_2|^2}{q_1^2}   \nonumber  
    \,,  
      W(k R)W^2(q_{2} R) \frac{k^2}{q_1^2} \nonumber 
    \,,  W^2(q_1 R)W^2(q_2 R) \Bigg\}\nonumber \,, \\
      &  M^{(4)}(\mathbf{k}-\mathbf{q}_1,\mathbf{q}_1,\mathbf{q}_2,-\mathbf{q}_2)  \overset{q_2\rightarrow\infty}{\propto}  \Bigg\{
    W(kR) \frac{k^2}{q_2^2} , W(kR)\frac{k^4}{q_2^4} , 
    \\
    & \,W(q_1R)W(|\mathbf{k}-\mathbf{q}_1| R)\frac{|\mathbf{k} - \mathbf{q}_1|^2}{q_2^2}, 
     \,W(q_1R)W(|\mathbf{k}-\mathbf{q}_1| R)\frac{ q_1^2}{q_2^2}, \nonumber  
     \\
    &W(kR)W^2(q_2 R)F^{(2)}(\mathbf{k}-\mathbf{q}_1,\mathbf{q}_1) \, ,  W(|\mathbf{k}-\mathbf{q}_1| R)W(q_1 R)W^2(q_2 R) \Bigg\} \,,\nonumber  
\end{eqnarray}
\begin{eqnarray}
    &M^{(5)}(\mathbf{k},\mathbf{q}_1,-\mathbf{q}_1,\mathbf{q}_2,-\mathbf{q}_2)  \overset{q_1\rightarrow\infty}{\propto}  \Bigg\{
    W(kR) \frac{k^2}{q_1^2} , W(kR)\frac{k^4}{q_1^4},
    \\
    &  W(|\mathbf{k}+\mathbf{q}_2|R)W(q_2 R)\frac{|\mathbf{k} + \mathbf{q}_2|^2}{q_1^2} \,  ,
    W(|\mathbf{k}+\mathbf{q}_2|R)W(q_2 R)F^{(2)}(\mathbf{q}_2, \mathbf{k})\frac{q_2^2}{q_1^2} , \nonumber  
    \\
    &  W^2(q_{1}R)W(kR)F^{(3)}(\mathbf{k}, \mathbf{q}_2, -\mathbf{q}_2) , W^2(q_{2}R)W(kR)\frac{ q_2^2}{q_1^2} \nonumber
    , W^2(q_{2}R)W(kR)\frac{k^2}{q_1^2},  \nonumber 
    \\
    &  W( |\mathbf{k}+\mathbf{q}_2| R)W(q_{2} R)W^2(q_{1} R) F^{(2)}(\mathbf{q}_2, \mathbf{k}) \, ,
     W^2(q_1 R)W^2(q_2 R)W(k R)\Bigg\} \nonumber\,.
\end{eqnarray} 
We can see that the kernels will contain very intricate dependence coming from the filtering. Many of those terms are exponentially suppressed by filters when integrated.
However, one of those terms in the $M^{(5)}$ kernel will remain in the single-hard limit:
\begin{eqnarray} \label{eq:M5newterm}
  M^{(5)}(\mathbf{k},\mathbf{q}_1,-\mathbf{q}_1,\mathbf{q}_2,-\mathbf{q}_2)  
  &\supset&
  W(|\mathbf{k}+\mathbf{q}_2|R)W(q_2 R)
  F^{(2)}(\mathbf{q}_2, \mathbf{k})F^{(3)}(\mathbf{q}_2, \mathbf{q}_1,-\mathbf{q}_1) \nonumber \\
  &\overset{q_1\rightarrow\infty}{\propto}&
  W(|\mathbf{k}+\mathbf{q}_2|R)W(q_2 R)F^{(2)}(\mathbf{q}_2, \mathbf{k})\frac{q_2^2}{q_1^2} \,.
\end{eqnarray}
This term comes from the bispectrum of $\delta$ and it is a drawback for bringing information from the three-point function down to the two-point function.
Therefore,  in order to cancel out the  $P_{A}^{(15)}$  cutoff dependence one needs to introduce at two loops a term like
\begin{gather} 
  P_{\rm bispec} (k) =  P_L(k)\int_\mathbf{q}  P_L(q)F^{(2)}(\mathbf{k},\mathbf{q}) \, q^2 \,.
 \end{gather}
 
 Also the set of free parameters needed to fix the large-scale limit at two loops is larger than at one loop. Due to the $F$ independent terms in the $M$ kernels, one needs to include those six terms: 
     \begin{equation} \label{eq:lowkcontrib2L}
        \{ k^0, k^2, k^4,P_L, P^{1L}, P_{\rm quad}\} \,.
    \end{equation}
Therefore, the two-loop full set of free parameters is
\begin{eqnarray}
    \{k^0, k^2, P_L, P^{1L}, P_{\rm quad}, k^2P_L, k^2P^{1L}, k^4,k^4P_L, k^2 P_{\rm quad},P_{\rm bispec} \} \,,
\end{eqnarray} 
which sums up to a total of eleven free terms. It more than doubles the number of free parameters of the two-loop calculation for $\delta$ and approaches the number of free-parameters of the three-loop calculation for $\delta$ \cite{Konstandin:2019bay}.

Here it is important to point out that, 
although there are many new free parameters to describe the power spectrum of $A$ in comparison to the one of $\delta$, several of these new parameters are degenerated with the ones present in the bias expansion Eq.~\eqref{eq:P1l_bias} (e.g., stochastic parameters). For instance, the number of free parameters is not increased when biased tracers are considered at one loop, as is the case in the analysis of galaxy surveys.

\bibliographystyle{JHEP}
\bibliography{main}

\end{document}